\documentclass[twocolumn,preprintnumbers,amsmath,amssymb]{revtex4}
\usepackage{graphicx}
\usepackage[colorlinks=true,urlcolor=blue,citecolor=blue,linkcolor=blue]{hyperref}
\usepackage{amsmath}
\usepackage{amssymb}
\usepackage{amsthm}
\usepackage{amsfonts}
\usepackage{enumerate}
\usepackage{subfigure}
\usepackage{centernot}
\usepackage{tikz}
\usepackage{ulem}

\usepackage{mathbbol}

\begin{document}

\title{Correlated helimagnetic configuration in a nonsymmorphic magnetic nodal semimetal}

	 \author{ Xi Luo$^{1}$}\thanks{These two authors contribute equally.}
     \author{ Yu-Ge Chen$^{2}$ }\thanks{These two authors contribute equally.}
     \author{Ye-Min Zhan$^{3}$}
 \author{ Yue Yu$^{3}$}
	\thanks{Correspondence to: yuyue@fudan.edu.cn}

 \affiliation { 
 	1. College of Science, University of Shanghai for Science and Technology, Shanghai 200093, China\\
    2. Institute for Quantum Science and Technology, Department of Physics, Shanghai University, Shanghai 200444, China\\
     3. State Key Laboratory of Surface Physics and Department of Physics, Fudan University, Shanghai 200433,China\\ }

 \date{\today}

 \begin{abstract}
 {Nonsymmorphic magnetic Weyl semimetal materials such as ReAlX (Re=rare earth, X=Si/Ge) provide a unique opportunity to explore the correlated phenomena between Weyl fermions and nontrivial magnetic configurations. To be specific, we study a lattice model in which the magnetic configuration is determined by the competition among  ferromagnetic (FM) interaction, the Dzyaloshinskii-Moriya interaction, and the Kondo coupling $K_0$ to the Weyl fermion. Both quantum and finite-temperature phase transitions between FM and correlated nesting helical configurations are found. Different from the uncorrelated helimagnet that decouples from the Weyl fermions, this correlated helimagnet induces a magnetic Brillouin zone with a $K_0$-dependent nesting in the band structure of the conduction electrons instead of the monopole-like Weyl cone. By measuring the current induced by the chiral magnetic effect on the conduction electron with nesting Weyl nodes, one can distinguish the correlated nesting helical order from the ferromagnetism because the chiral magnetic effect is considerably suppressed in the former case. These properties we find here may explain the experimental observations in ReAlX.}
 \end{abstract}
\maketitle

\section{{Introduction}} 
{Recently, the nodal semimetals in a set of nonsymmorphic magnetic materials, ReAlX(Re=rare earth, X=Si/Ge)}, attract widespread attention  \cite{Xu2017,Chang2018,Sanchez2020,Destraz2020} and a type of Weyl semimetals with breaking of both the time-reversal $\Theta$ and inversion $P$ symmetries was found in ReAlX \cite{Tafti2023,Tafti2021,Tafti2020,Broholm2021,Li2023,Alam2023,Yu2023,Yang2023,Terashima2024,Yamada2024}. These magnetic materials belong to the body-centered tetragonal crystal system with a non-centrosymmetric  space group $I4_1md$. Due to the $P$ breaking, the Weyl nodes possess different energies. Meanwhile, the $f$-orbital electrons of the rare earth elements provide a strong magnetic moment, so that a strong Ruderman-Kittel-Kasuya-Yosida ({RKKY}) interaction can exist by mediating Weyl fermions.  The {RKKY} interaction induces competition among the Heisenberg, Kitaev-Ising, and Dzyaloshinskii-Moriya (DM) interactions \cite{Xiao2015,rkky2,ZhouPRB2015,ZhouPRB2017,LL2025}, which leads to the emergence of several nontrivial magnetic configurations, such as spiral, helical, hedgehog, skyrmion, and meron-antimeron pairs \cite{Tretiakov2021}. 

 The ReAlX materials provide a unique opportunity to study the correlated phenomena that come from the interplay between Weyl fermions and nontrivial magnetic configurations through the Kondo coupling \cite{FritzPRB2015}, e.g., new exotic phenomena such as the topological Hall effect and room temperature anomalous Hall effect \cite{Tafti2023,Li2023}. This raises the possibility of preparing high-density ultra-fast storage devices with these materials \cite{Beidenkopf2022,BatistaPRB2021,Gupta2021,YouAPL2021,Incorvia2023}.

Physically, those nontrivial magnetic orders rotate the Weyl fermion's spin configuration in real space via the Kondo coupling, which competes with the monopole topology of the Weyl nodes in momentum space. This competition may further induce new   {correlated} phenomena and correlated topological states of matter that lack clear explanations thus far.

 In this {paper}, we focus on the correlated nesting helical magnetic configuration (CNHMC), which has been observed in ReAlX materials. For example, since the ratio between out-of-plane and in-plane magnetic susceptibility $\chi_c/\chi_a$ is very close in SmAlSi at $2K$, a helimagnet with ${\bf Q}=\frac{2\pi}{3}(0.979, 0.979, 0)$  emerges, which is consistent with the nesting vector between Weyl nodes \cite{Tafti2023}. In CeAlGe, a phase transition from the canted magnetic order to the helimagnet was observed at $4.5K$. The helimagnet also provides a nesting vector between the Weyl nodes and stabilizes magnetic fluctuations from the magnetic transition temperature to the Curie temperature \cite{Yu2023}.  This phase transition to the helimagnet and the CNHMC effect are also reported for PrAlSi \cite{Tafti2020}, NdAlSi \cite{Broholm2021,Yamada2024}, CeAlSi \cite{Tafti2021}, and NdAlGe \cite{Yang2023}. 

  However, the mechanism of the magnetic phase transition and the origin of the emergence of the CNHMC in those materials remain unclear. 
  The specific mechanism behind the anomalous transport phenomena observed near $T=12.8$K in CeAlGe \cite{Yu2023}, such as the nearly zero magnetoresistance, the switch from negative to positive magnetoresistance, and the maximal suppression of thermal conductivity by magnetic fields is not yet fully understood.  {In Ref. \cite{Yu2023}, in which two of the present authors participated, a phenomenological analysis was conducted by treating the magnetic order and Weyl fermions separately (termed uncorrelated Weyl node nesting), while the CNHMC effects and magnetic phase transition remain unexplored.}

 Conventionally, the magnetic configuration ${\bf M}$ and Weyl fermions are separately studied.  However, in ReAlX, the strong Kondo coupling plays an important role, so this separated treatment is not valid.  {In this {paper}, we develop an exact solution to the {Kondo coupling} problem {in the continuum limit}, identifying a characteristic zero-energy mode at the Fermi surface associated with helical magnetic ordering.}  We then take this correlated ${\bf M}$ configuration as a background field to deal with the conduction electrons. When we determine the phase diagram of the correlated magnetism, we fix the fermion spin configuration. With these approximations, we see the phase transitions from  an ordered phase, e.g., the ferromagnetic (FM) one, to the CNHMC both at zero temperature and finite temperature. We calculate the current induced by the chiral magnetic effect (CME) both in the FM phase and the CNHMC and find that the current for the latter is an order of magnitude less than that for the former. This may explain the nearly zero magnetoresistance  near $T=12.8$K in CeAlGe \cite{Yu2023}.
 
 {The rest of this paper is organized as follows. In Sec. \ref{secii}, we construct a minimal two-band model for a Weyl semimetal that breaks both time-reversal and inversion symmetries and  is coupled to the magnetic configuration ${\bf M}$. We also consider the continuum limit of the model with the helical magnetic configuration, and an exact solution is found at zero energy. In Sec. \ref{seciii}, we draw the band structures of the lattice model with CNHMC and find that the band structure forms a magnetic Brillouin zone, and the Weyl nodes are folded onto each other for small Kondo coupling. For large Kondo coupling, the Weyl nodes move towards each other, eventually annihilating and opening a gap. In this section, we also consider the magnetic phase diagram by calculating the ground state energy per site at zero temperature and the free energy per site at finite temperature. We find a competition between the ferromagnetic phase and the helical one, which is consistent with the experiments. In Sec. \ref{seciv}, we calculate the CME of the lattice model. We find that the corresponding magnetoresistance of CNHMC is strongly suppressed compared to that of the ferromagnetic case, which may relate to the nearly zero magnetoresistance observed in CeAlGe \cite{Yu2023}. The final part is devoted to conclusions. }
 
 %\textcolor{red}{check the citation of appendices}
 
\section{{Lattice model and continuum limit}}  \label{secii}
\subsection{Lattice model}
 The common model study for the materials ReAlX starts from the calculation of the band structure of this body-centered tetragonal crystal system with a non-centrosymmetric space group $I4_1md(109)$. The Weyl nodes are found near the Fermi level and are coupled to the local moment from the $f$ electron in the rare-earth element. Such a strategy is fine but it is  difficult to understand the basic physics due to the complicated numerics. 

  {Due to the crystal symmetry and observed large anomalous Hall effects  \cite{Tafti2020,Tafti2021,Yang2023,Li2023,Terashima2024}, the conduction electrons in ReAlX are Weyl semimetals with both $\Theta$ and $P$ symmetry breaking.}  {Remarkably, ReAlX-type materials exhibit dramatic changes in transport properties (e.g., magnetoresistance) across the phase transition to the helimagnetically ordered state.} To see the basic physics behind the numerics, we study a simplified cubic lattice model in which the conduction electrons are Weyl semimetals without $\Theta$ and $P$ symmetries and have two Weyl nodes that are adjusted to be near the Fermi level.  The electrons couple to  ${\bf M}_i$ via the Kondo coupling while the nearest neighbor ${\bf M}_i$ and ${\bf M}_j$ are coupled through the FM Heisenberg and the DM interactions:
\begin{eqnarray}
H_{~~~}&=&H_W+H_K+H_H+H_{DM},\label{htotal}\\
    H_{W~}&=&	\sum_j[(-t_xc_j^\dagger \sigma_z c_{j+\hat{x}}-t_yc_j^\dagger \sigma_z c_{j+\hat{y}}-t_zc_j^\dagger \sigma_z c_{j+\hat{z}}\nonumber\\
	&&-it'_xc_j^\dagger\sigma_xc_{j+\hat{x}}
    -it'_yc_j^\dagger\sigma_yc_{j+\hat{y}}-iv_0c_j^\dagger\sigma_0c_{j+\hat{z}}\nonumber\\
    &&+h.c.)
    +mc_j^\dagger \sigma_zc_j],\nonumber\\
    H_{K~~}&=&\sum_jK_0{\bf {s}}_j\cdot {\bf M}_j, \quad H_H=\sum_{\langle ij\rangle}J_{0}{\bf M}_i\cdot {\bf M}_j,\nonumber\\
    H_{DM}&=&\sum_{\langle ij\rangle}{\bf D}_0\cdot {\bf M}_i\times {\bf M}_j. \nonumber
\end{eqnarray}
For simplicity,  we restrict $H_W$ to be a minimal two-band model  \cite{feiye2016,Trivedi2017} and the Weyl nodes are located at $(0,0,\pm k_0)$ with $\cos k_0=(m/2-t_x-t_y)/t_z$. $c_j^\dag$ (the spin index  omitted) is the creation operator of the conduction electron at site $j$ with spin ${\bf s}_j=\frac{1}{2}c_{j}^\dag {\boldsymbol{\sigma} } c_{j}$.  For the FM case, $J_0<0$. In the following, we use   {$t_x=t_y=t_z=t'_x=t'_y=t=1,m=5t$ and $k_0=\pi/3$. The chemical potential has been set to zero}. $H_K$ is the  on-site Kondo coupling between ${\bf s}_i$ and ${\bf M}_i$ with the coupling strength $K_0$. $H_H$ and $H_{DM}$ are the nearest-neighbor FM Heisenberg and DM interactions, respectively,  {both originating from RKKY coupling between Weyl fermions and rare-earth local moments}.  {Although  our simplified model adopts a cubic lattice, we can map it to the tetragonal  symmetry $I4_1md$ through parameter tuning. Specifically, tuning $t_x=t_y\neq t_z$ and $t_x'=t_y'$ and substituting $k_z\rightarrow \frac{a}{c}k_z$ (where $a \neq c$ are the $xy$-plane and $z$-direction lattice constants) reproduces the ReAlX structure. The typical parameters in ReAlX are $a\sim 4 $ $\mathring{A}$, $c\sim 14$ $\mathring{A}$, $t_{x,y}\sim 100$ meV, and $t_z\sim 70$ meV \cite{Tafti2020,Broholm2021,Yamada2024,Tafti2021,Yu2023,Tafti2023,Yang2023}. For clarity in studying CNHMC's topological properties, we use an isotropic model.}  {Our simulations demonstrate that CNHMC-induced band folding significantly suppresses the CME, while reproducing the experimentally observed magnetoresistance trends in ReAlX compounds.}

 {Beyond the helical magnetic order, other complex magnetic configurations (e.g., canted orders) and physical phenomena emerge due to multiple Weyl node pairs and lattice non-symmorphic features, though these are beyond our present scope. Our work concentrates on the ferromagnetic-to-helical phase transition and its topological characteristics, while a four-band model addressing multiple Weyl pairs will be developed in future studies.}

 When solving the conduction electron problem, we use the adiabatic approximation, i.e., assuming the conduction electrons are fast-moving with respect to the fluctuations of ${\bf M}$ when ${\bf M}$ is considered a background. On the other hand, the configuration of $\bf M$ is determined by the competition of the variational ground state energies when the conduction electron spin is assumed to be polarized by ${\bf M}$ 's configuration, the CNHMC or ferromagnet.     

For the helical magnetic configuration, without loss of generality, one takes
${\bf M}=(\sin [Qz],\cos [Qz],0).$
The Hamiltonian for Weyl fermions with the Kondo coupling becomes
\begin{eqnarray}
    H_0&=&H_W+H_K\label{h0}\\
   &=&\sum_{{\bf k},\sigma}[ (m-2t_x\cos k_x-2t_y\cos k_y\nonumber\\
   &-&2t_z\cos k_z)c_\sigma^\dag({\bf k})\sigma_zc_\sigma({\bf k})+2t'_x \sin k_x c_\sigma^\dag({\bf k}) \sigma_x c_\sigma({\bf k}) \nonumber\\ 
    &+&2t_y' \sin k_y c_\sigma^\dag({\bf k})\sigma_yc_\sigma({\bf k}) +2v_0\sin k_zc_\sigma^\dag({\bf k})\sigma_0c_\sigma({\bf k})\nonumber\\
  &+&(c^\dagger_{\uparrow}(k_x,k_y,Q+k_z)(-i\frac{K_0}{2})c_\downarrow(k_x,k_y,k_z)+h.c.)],\nonumber 
\end{eqnarray}
where $\uparrow$ and $\downarrow$ are the spin indices. The last term in Eq. (\ref{h0}) is the Kondo coupling between the Weyl fermion and the helical magnetic configuration.  Notice that the momentum shift from $k_z$ to $k_z\pm Q$ comes from the helical wave number $Q$.

\subsection{{Continuum limit and CNHMC effects }}   
\begin{figure}
\includegraphics[width=0.23\textwidth]{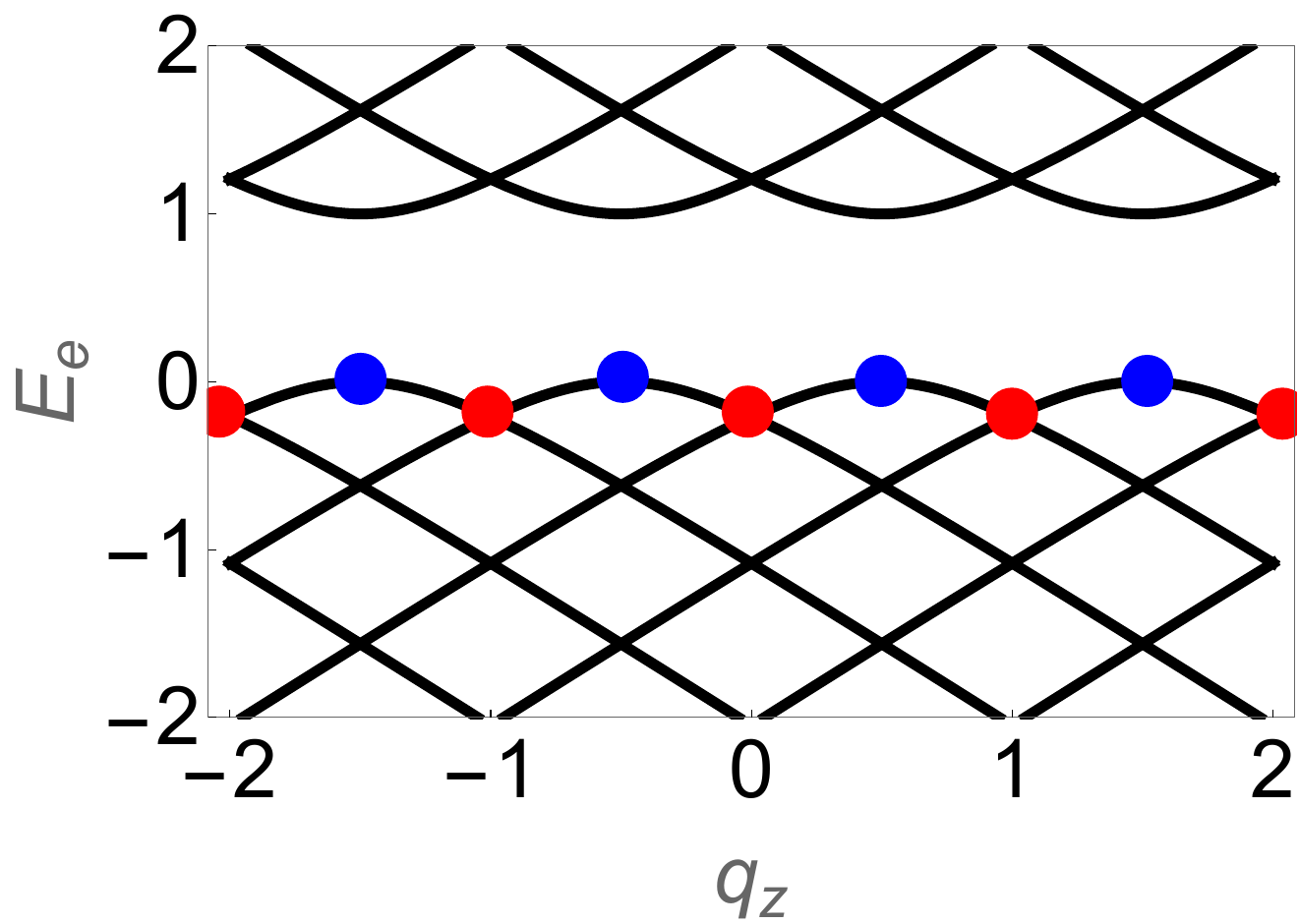}
\caption{(color online)    {The energy spectrum in the continuum limit for Weyl fermions coupled with a helimagnet along the $q_z$-direction. We set $Q=K_0=1$. The helimagnet  introduces a magnetic Brillouin zone. The red points stand for the Weyl nodes and the blue points stand for the  zero modes (see  {Appendix \ref{app1}}).}}
	\label{fig0}
\end{figure}

{To demonstrate the energetic preference for helical ordering in Kondo-coupled Weyl systems, we first establish its stability in the continuum limit [Eq. (\ref{h0})] before validating with lattice simulations, and demonstrate that the correlated helimagnet constitutes an allowed ground-state configuration.} When $|Q|\ll 2|k_0|$, due to the energy offset $b_0$ between the two Weyl nodes in a non-centrosymmetric system, we can integrate {out one Weyl fermion that is below the chemical potential,}  say,  {$\psi_-$}. The effective equations of motion of ${\bf M}$ and the conduction electron $\psi=\psi_+$ at a given Weyl valley in the continuum limit are given by \cite{Yu16}  
\begin{eqnarray}
	&&-i\boldsymbol\sigma\cdot(\nabla+i\frac{K_0}2{\bf M})\psi=E_e\psi,\nonumber\\
	&&K_0{\bf s}+\frac{D}{2}\nabla\times {\bf M}-{J}\nabla^2{\bf M}=0, \label{EOM}
\end{eqnarray}
where $J\propto K_0^2-2J_0$ and   $D\propto v_- K_0^2+4D_0$ \cite{Xiao2015,rkky2}.  
In the correlated case, the helimagnet  remains an exact $E_e=0$ solution  with 
    $Q=\frac{D\pm \sqrt{D^2+16|J|K_0}}{4|J|},$
which reduces to the uncorrelated helimagnet or the ferromagnet when $K_0=0$ i.e.,  $
   Q= |{\bf Q}|=|D/2J|$ and/or $Q=0$
 \cite{Nagaosa2012}.   When $K_0\ne 0$,  ${\bf M}$ and $\psi$ are correlated via the Kondo coupling.  The energy spectrum of the conduction electron $\psi$ is shown in Fig. \ref{fig0}. The zero mode solutions are localized at the boundaries of the magnetic Brillouin zone which is introduced due to the helimagnet (for details, see  {Appendix \ref{app1}}).

 {When ${\bf Q}$ of the helimagnet forms a commensurate nesting vector between the Weyl nodes, the helimagnet becomes a CNHMC. This nesting effect has been used to explain the experimental results in CeAlGe \cite{Yu2023}, but the uncorrelated helimagnet state is not available in ReAlX.  By considering the CNHMC, the low-energy effective Hamiltonian of the conduction electron reads
 $H_{eff}'=\sum_{\bf k}\Phi^\dag_{\bf k}h'_{\bf k}\Phi_{\bf k}$,
 where
\begin{eqnarray}  
    h'_{\bf k}&=&k'_x\sigma_x \tau_0+k'_y\sigma_y\tau_0+k'_z\sigma_z\tau_z\nonumber\\
&+&a_0k_z'\sigma_0\tau_0+b_0\sigma_0\tau_z+K_0\sigma_y\tau_x/2, \label{h'}
\end{eqnarray}
and  $\Phi=(c_{+\uparrow},c_{+\downarrow},c_{-\uparrow},c_{-\downarrow})$ with $\pm$ labeling two Weyl valleys;  $\sigma$ and $\tau$ are the matrices that act on spin and valley indices, respectively. $k_x'=2t_x'k_x$, $k_y'=2t_y'k_y$, $k_z'=2t_z(\sin{k_0})k_z$, $a_0={v_0(\cot{k_0})}/{t}$, and $b_0=2v_0(\sin{k_0})$. The spectra of Hamiltonian (\ref{h'}) read
\begin{eqnarray}
&&E_{\pm,\pm}=a_0k'_z\pm\nonumber\\
&&\sqrt{{\bf k}'^2+(K_0/2)^2+b_0^2\pm\sqrt{K_0^2(k_y'^2+k_z'^2)+4b_0^2{\bf k}'^2}}.
\end{eqnarray}
In the $K_0\rightarrow 0$ limit, $b_0$ becomes the energy offset between the two Weyl nodes, and $a_0$ controls the type of the Weyl node. {When $b_0\rightarrow 0$, the CNHMC does not open a gap but drives the Weyl semimetal to a nodal-ring semimetal when $k_y'^2+k_z'^2=K_0^2/4$.} 
}

\section{Band structure and phase diagram} \label{seciii}
 \subsection{{Band structures of the fermion}}  
 {We now turn back to the lattice model.} Similar to dealing with integer quantum Hall effects with the lattice model  \cite{Kohmoto1989,Kohmoto1990,Hatsugai1993} (also see  {Appendix \ref{app2}} for a brief review), one  defines
$|\Psi\rangle=\sum_{j=1}^q\psi_jc^\dag (k_x,k_y,k_z^0+Q j)|0\rangle,$
where $Q=2\pi p/q$ with $p$ and $q$ coprime,  $k_z=k_z^0+Q j$, and $\psi_{j+q}=\psi_j$. Then, 
 the Schr\"odinger equation $H_0|\Psi\rangle=E|\Psi\rangle$ is reduced to the Harper equations 
\begin{eqnarray}
   &[&m-2t_x\cos k_x-2t_y\cos k_y-2t_z\cos (k_z^0+Qj)]\sigma_z\psi_j\nonumber\\
    &+&2t_x' \sin k_x   \sigma_x \psi_j+2t_y' \sin k_y \sigma_y\psi_j+2v_0 \sin(k_z^0+Qj)\psi_j\nonumber\\
    &+&(i\frac{K_0}{2})((\frac{\sigma_z-1}{2})\psi_{j-1}+(\frac{\sigma_z+1}{2})\psi_{j+1})\nonumber\\
    &=&E(k_x,k_y,k_z^0)\psi_j. \label{para}
\end{eqnarray}
The Harper equations have $q$ eigenvalues for a given $(k_x,k_y,k_z^0)$. The original band is split into $q$ subbands due to the Kondo coupling to the CNHMC, and each subband has a reduced magnetic Brillouin zone: $-\pi\leq k_x, k_y\leq \pi$ and $-\pi/q \leq k_z^0\leq \pi/q$. 

\begin{figure}
	\begin{minipage}{0.23\textwidth}	\centerline{\includegraphics[width=1\textwidth]{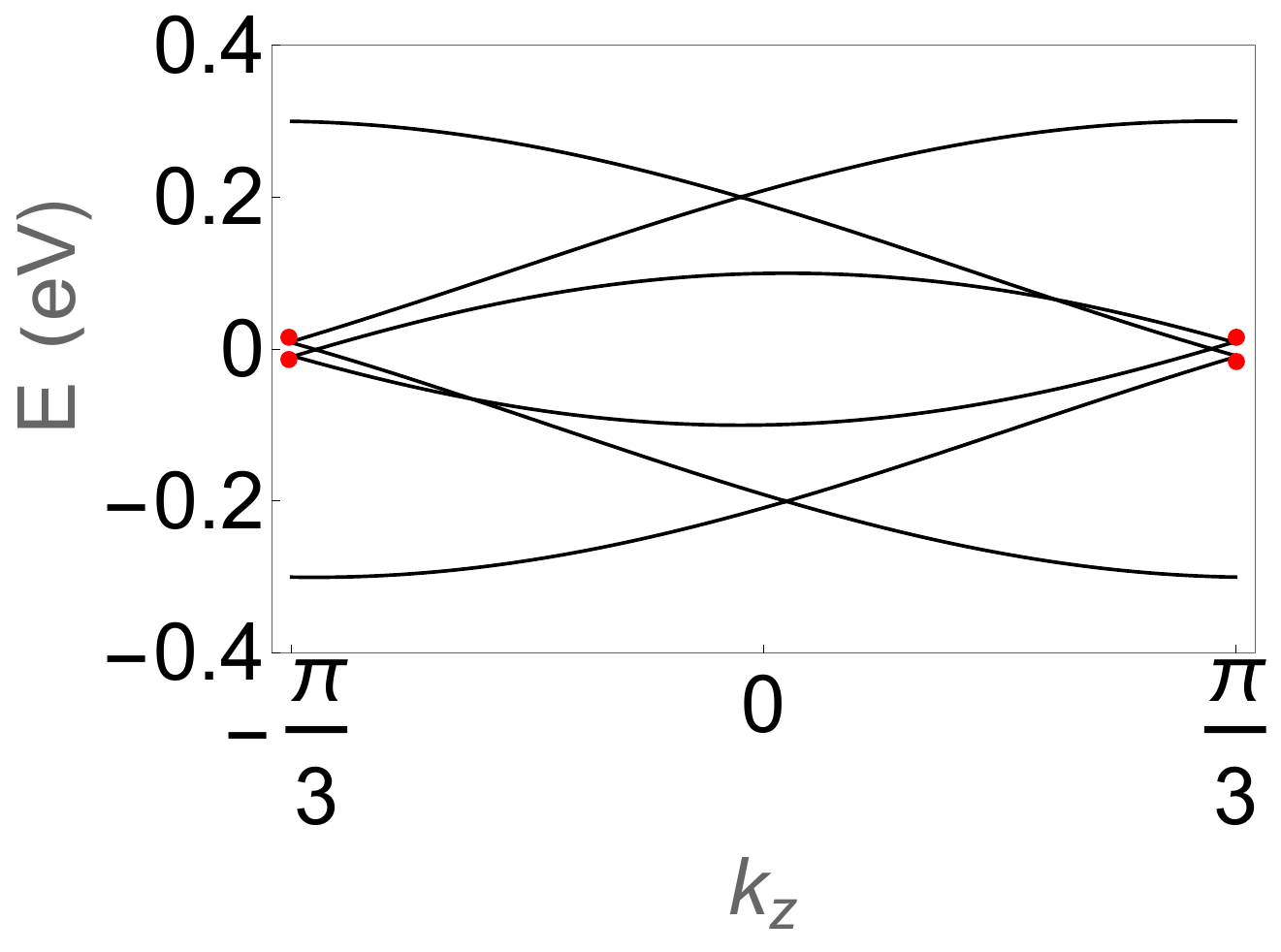}}
		\centerline{(a) }
	\end{minipage}
	\begin{minipage}{0.23\textwidth}		\centerline{\includegraphics[width=1\textwidth]{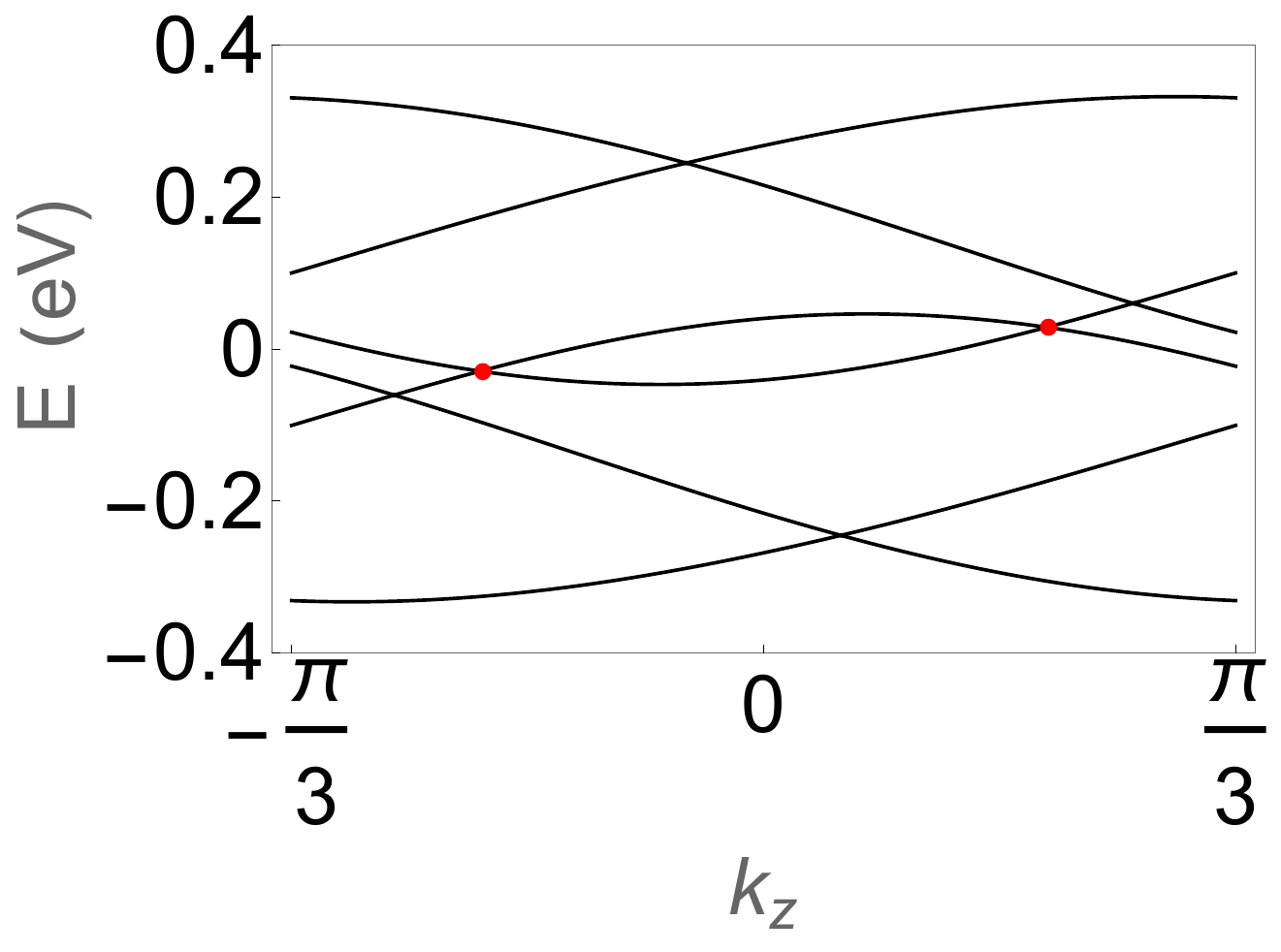}}
		\centerline{(b) }
	\end{minipage}
    	\begin{minipage}{0.23\textwidth}		\centerline{\includegraphics[width=1\textwidth]{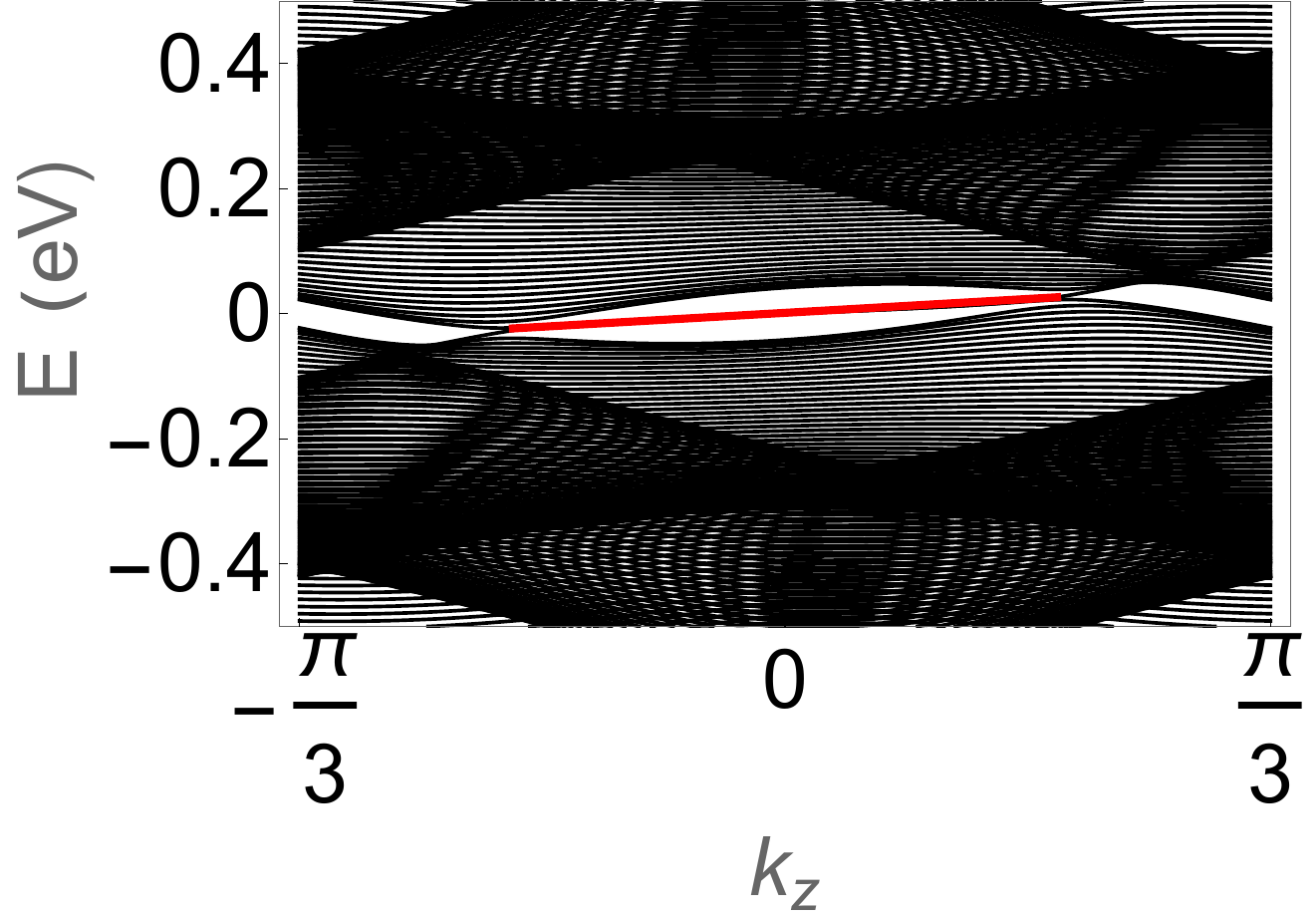}}
		\centerline{(c) }
	\end{minipage}
        	\begin{minipage}{0.23\textwidth}	\centerline{\includegraphics[width=1\textwidth]{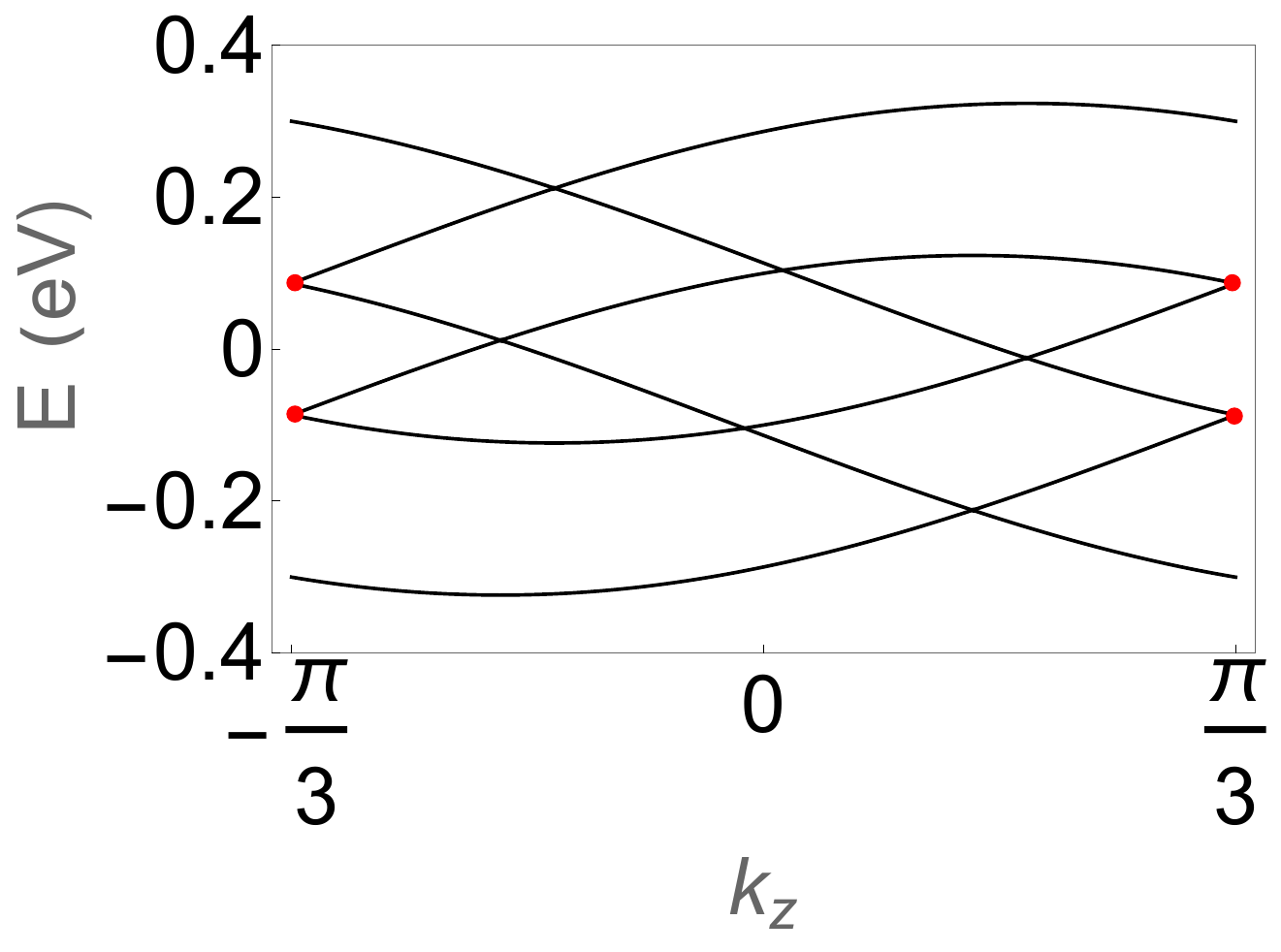}}
		\centerline{(d) }
	\end{minipage}
\caption{(color online)
 {The band structures of Weyl fermions with CNHMC for parameters $Q=2\pi/3$, $t=t'=0.1$ eV, and $k_x=k_y=0$.  (a) $K_0=10$ meV. (b) $K_0=200$ meV. (c)   Open boundary conditions for (b) with 50 sites along the $y$ axis. (d) $K_0=10$ meV. Parameter values: (a-c) $v_0=5$ meV  and (d) $v_0=50$ meV. The red points in (a), (b), and (d) indicate the Weyl nodes.  The red line in (c) represents the Fermi arc. }}
	\label{fig1}	
\end{figure}

In ReAlX, the Fermi velocity is $\sim1eV\mathring{A}$, and the lattice constant is $\sim10\mathring{A}$, and then $t\sim 100meV$. The  Curie temperature is about $30K$, and then $|J_0|\sim 2.5meV$.  From the first-principles calculations, $|K_0|\sim |D_0|\sim |J_0|$ are of the same order \cite{Tafti2023,Tafti2021,Tafti2020,Broholm2021,Yamada2024,Li2023,Alam2023,Yu2023,Yang2023,Terashima2024}. With these parameters, we plot the band structures of $H_0$ (\ref{h0}) with the CNHMC in Fig. \ref{fig1}.  {The observed helimagnetic ordering in SmAlSi with nesting vector ${\bf Q}=\frac{2\pi}{3}(0.979, 0.979, 0)$ \cite{Tafti2023} motivates our choice of the commensurate $Q=2\pi/3$ ($1/3$ Brillouin zone length)  as the Weyl node nesting vector.} We find that, for the commensurate nesting $Q=2\pi/3$, the Weyl nodes are folded to the  boundaries of the magnetic Brillouin  zone for a small  $K_0$ (see Fig. \ref{fig1}a).  {The monopole charges of the left and right Weyl nodes (red points) in Fig. \ref{fig1}b are $-1$ and $+1$, whereas the monopole charges of the other band crossings induced by the band folding of CNHMC  are zero. This topological feature is also confirmed by Fermi arc states under open-boundary conditions  in Fig. \ref{fig1}c.} As $K_0$ increases, the Weyl nodes move towards each other (see Fig. \ref{fig1}b), which may explain the slightly incommensurate behavior of the helimagnetic configuration observed in experiments \cite{Tafti2023}.    {For a strong enough $K_0$, the Weyl nodes merge together and finally open a gap. The Fermi arcs that connect the Weyl nodes will also evolve from shortening to disappearing; see Fig. \ref{fig1}c (for more details, see  {Appendix \ref{app3}}).} These phenomena reflect the competition between the monopole topology of the Weyl node in momentum space and the trivial topology of the helimagnet in real space.  The $P$-breaking term $v_0$  shifts the energy of every Weyl node and does not alter the location of the Weyl nodes (see Fig. \ref{fig1}d).

\subsection{Magnetic phase diagram} 
We turn to study the magnetic phase diagrams of the model Hamiltonian (\ref{htotal}). For a fixed magnetic configuration, which is either the ferromagnet or the CNHMC,  we compare the ground state energy per site at zero temperature and the free energy $f$ per site at finite temperature of  Eq. (\ref{htotal}). For the CNHMC, one takes $Q=2\pi/3$ and the ferromagnet is given by $Q=0$.  {To obtain the spectra, we use exact diagonalization  in the calculations and choose periodic boundary conditions for a cubic lattice with $L_x=L_y=L_z=20$. }

\begin{figure}
	\begin{minipage}{0.218\textwidth}	\centerline{\includegraphics[width=1\textwidth]{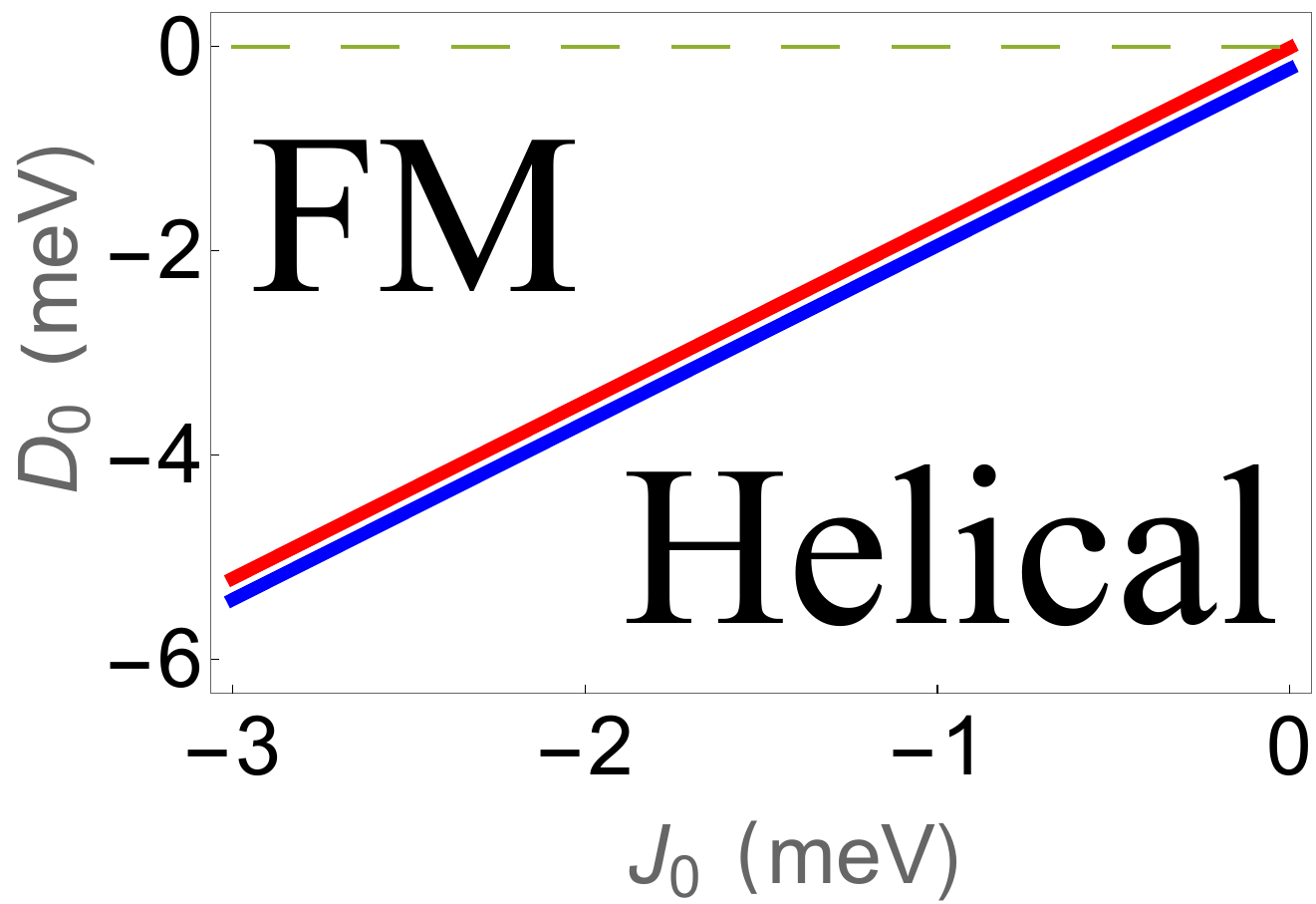}}
		\centerline{(a) }
	\end{minipage}
	\begin{minipage}{0.255\textwidth}		\centerline{\includegraphics[width=1\textwidth]{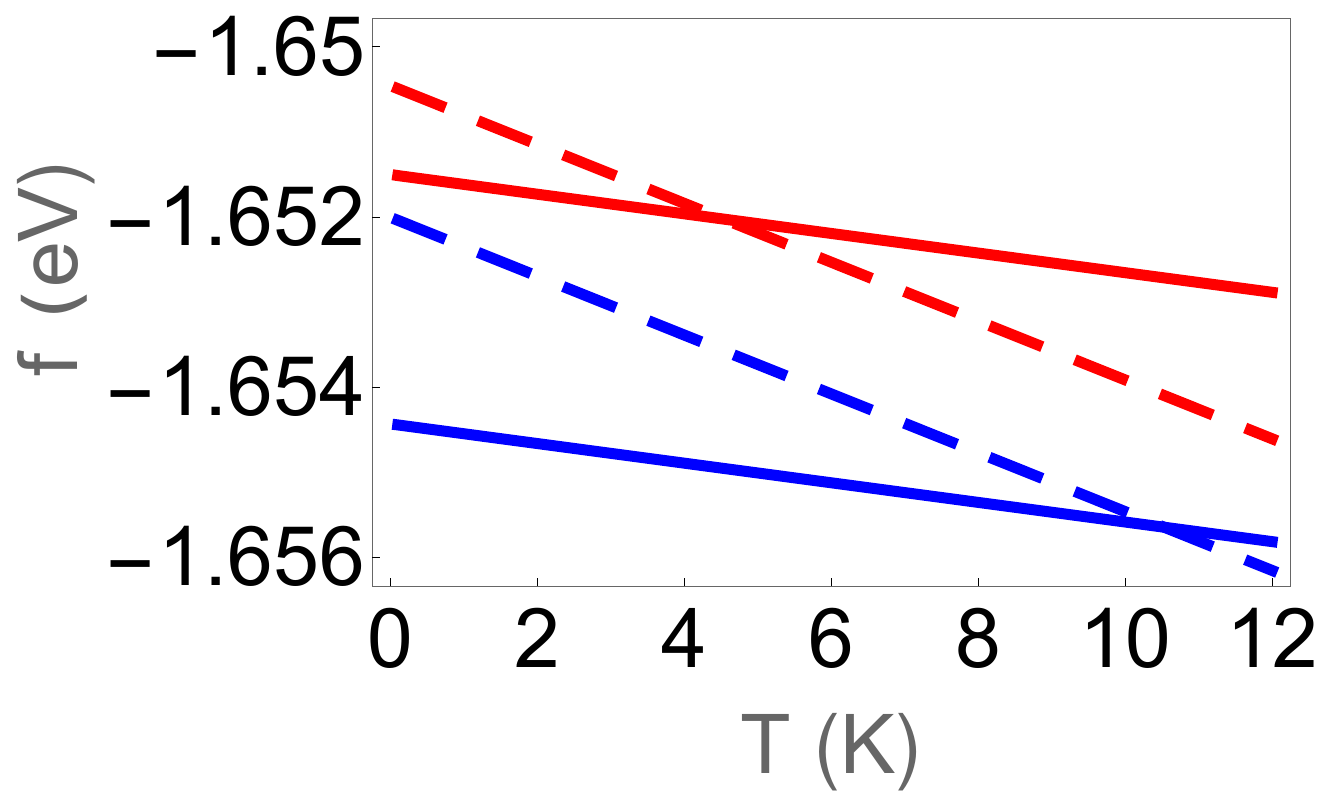}}
		\centerline{(b) }
	\end{minipage}
\caption{(color online)   {   {The parameters are the same as in Fig. \ref{fig1}a except for $K_0$. $K_0=10meV$ for the red lines and $K_0=50meV$ for the blue lines. (a) The zero-temperature phase diagram with $J_0$ vs $D_0$. The gray dashed line stands for the case with $D_0=0$, for which the ground state energy per site is the same for both phases along the lines. Above (below) the lines, the FM (helical) configuration is energetically favored.} (b) The free energy per site $f$ vs  temperature $T$ for $J_0=-3meV$ and $D_0=-4meV$. The solid (dashed) lines stand for $f$ with the FM (helical) configuration. 	}}
	\label{fig2}	
\end{figure}

At zero temperature, we find that for a fixed $J_0$, the CNHMC is favored for large  DM interaction $D_0$.   {The phase boundary moves downward when  $K_0$ increases (see Fig. \ref{fig2}a).} At finite temperature, by choosing $J_0=-3meV$, $D_0=-4meV$,  $K_0=10meV$,   {with the other parameters the same as those in Fig. \ref{fig1}a}, a phase transition occurs at about $T_c=4.5K$ (Fig. \ref{fig2}b), which is qualitatively consistent with experiments in Ref. \cite{Yu2023},  although the ordered phase is a canted magnetic order instead of the ferromagnet.   By increasing $K_0$, $T_c$ also increases (see Fig. \ref{fig2}b).

\section{{CME with CNHMC}} \label{seciv}
To further explore the effects of the CNHMC in this type of magnetic Weyl semimetals, we study the CME \cite{CME}. For an isolated Weyl node given by the low-energy Hamiltonian $h_w=b_0+ {\boldsymbol{\sigma}}\cdot({\bf k-b})$,  where ${\bf b}$ and $b_0$ denote the shift in momentum and energy, respectively. Its response to the external magnetic field is described by the topological $\theta$ term, $\theta_{{\bf r},t}{\bf E\cdot B}$, with $\theta_{{\bf r},t}=2({\bf b\cdot r}-b_0t)$ \cite{Burkov2012,Yamamoto2012,Tewari2013,Grushin2012}, which originates from the chiral anomaly of the Weyl fermion \cite{Adler1969,Jackiw1969,no-go} and corresponds to the charge density $\rho \propto {\bf b\cdot B}$ and current ${\bf j}\propto ({\bf b \times E}-b_0{\bf B})$. The ${\bf b}$-related terms contribute to the anomalous Hall effect, while the $b_0$ term contributes to the CME which is related to the giant negative magnetoresistance \cite{Franz2013,Burkov2015,Spivak2013,chen2015}.  In our model Hamiltonian  (\ref{h0}), $b_0\propto v_0$ and ${\bf b} \propto (0,K_0,0)$ in the FM limit $Q=0$. For the CNHMC with $|Q|=2|k_0|$, ${\bf b}$ is not well defined   {for small $K_0$} because the Weyl nodes are close to each other due  to  band folding.  

To explore the CME with {the CNHMC}, we apply a uniform magnetic field in the $z$ direction in the Hamiltonian (\ref{h0}). In the $x$-$y$ plane, the lattice sites are labeled by $(n, m)$, and the gauge $\theta^x_{(n,m),(n+1,m)}=0$ and $\theta^y_{(n,m),(n,m+1)}=2\pi \phi n$ is used \cite{Kohmoto1989,Kohmoto1990,Hatsugai1993}. 
Under this gauge, the flux through a unit plaquette is $2\pi \phi$. The current arising from the CME reads \cite{Franz2013}
\begin{equation}
J_z=\sum_{n,k_y,k_z}\langle\phi_{n,k_y,k_z}|\frac{\partial H_0(k_z)}{\partial k_z}|\phi_{n,k_y,k_z}\rangle n_F(E_n(k_y,k_z)),
\end{equation}
where $n_F$ is the Fermi distribution function. The periodic boundary conditions in the $y$ and $z$ directions are used. Both periodic and open boundary conditions are considered in the $x$ direction with a length $L_x$. We set $\phi=1/5$.

\begin{figure}
	\begin{minipage}{0.218\textwidth}	\centerline{\includegraphics[width=1\textwidth]{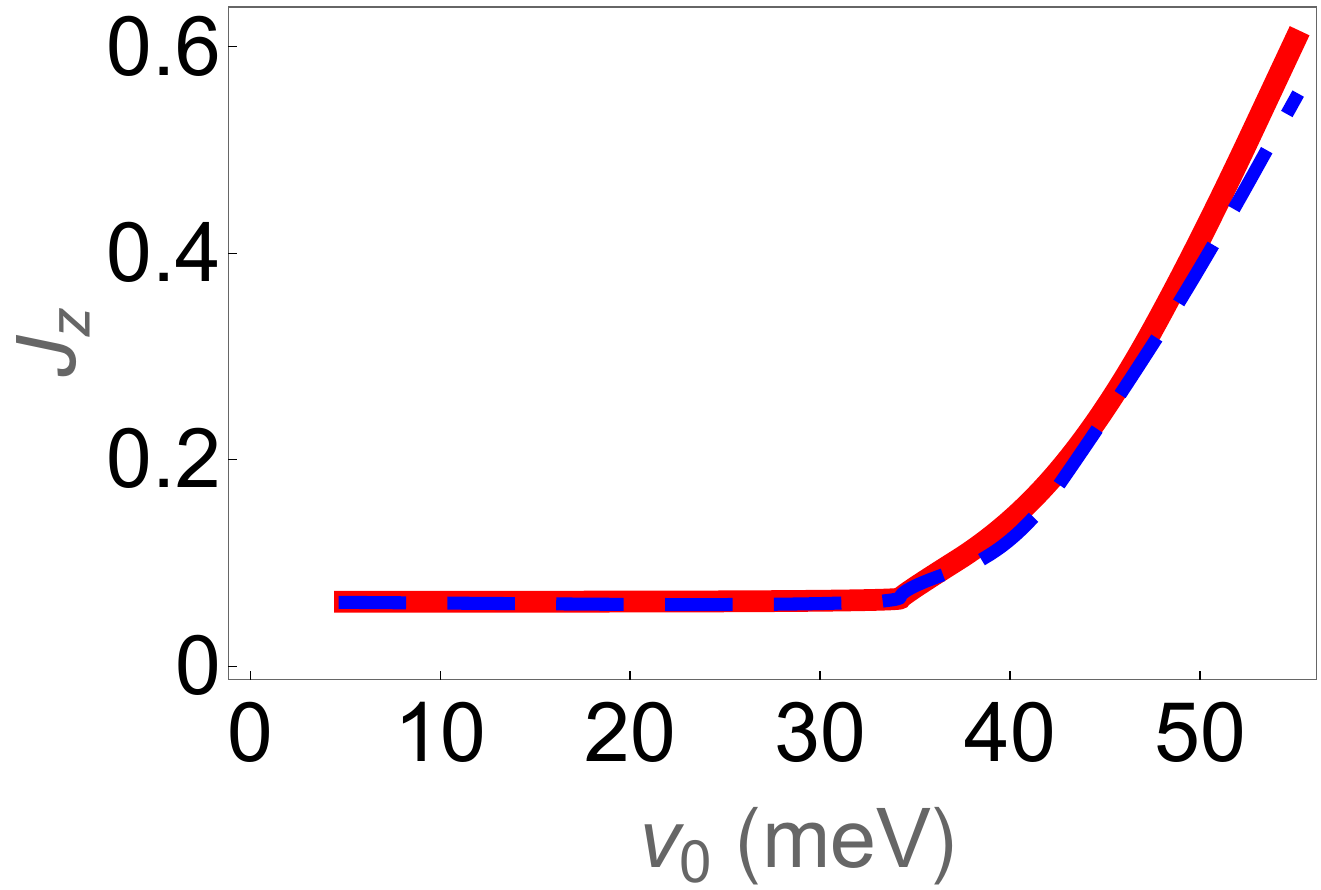}}
		\centerline{(a) }
	\end{minipage}
	\begin{minipage}{0.255\textwidth}		\centerline{\includegraphics[width=1\textwidth]{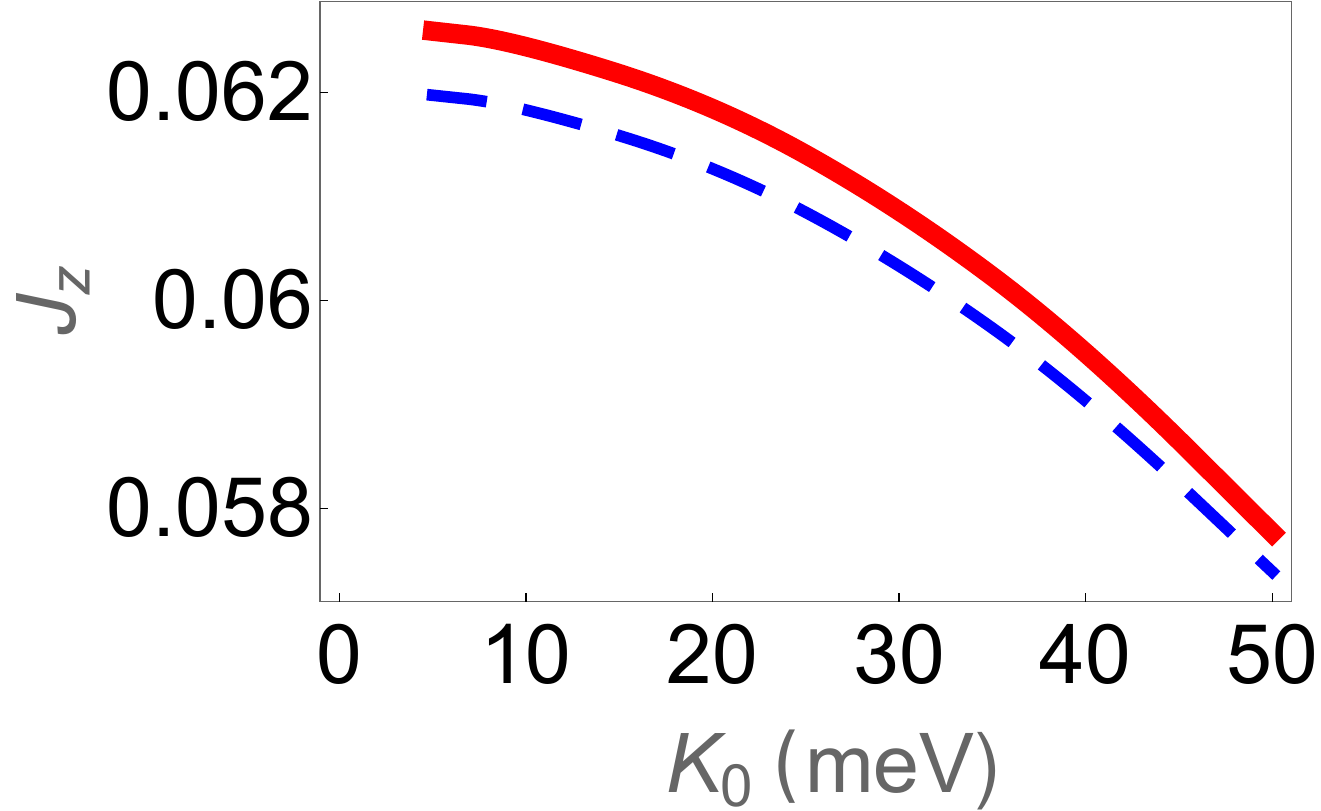}}
		\centerline{(b) }
	\end{minipage}
    \begin{minipage}{0.218\textwidth}	\centerline{\includegraphics[width=1\textwidth]{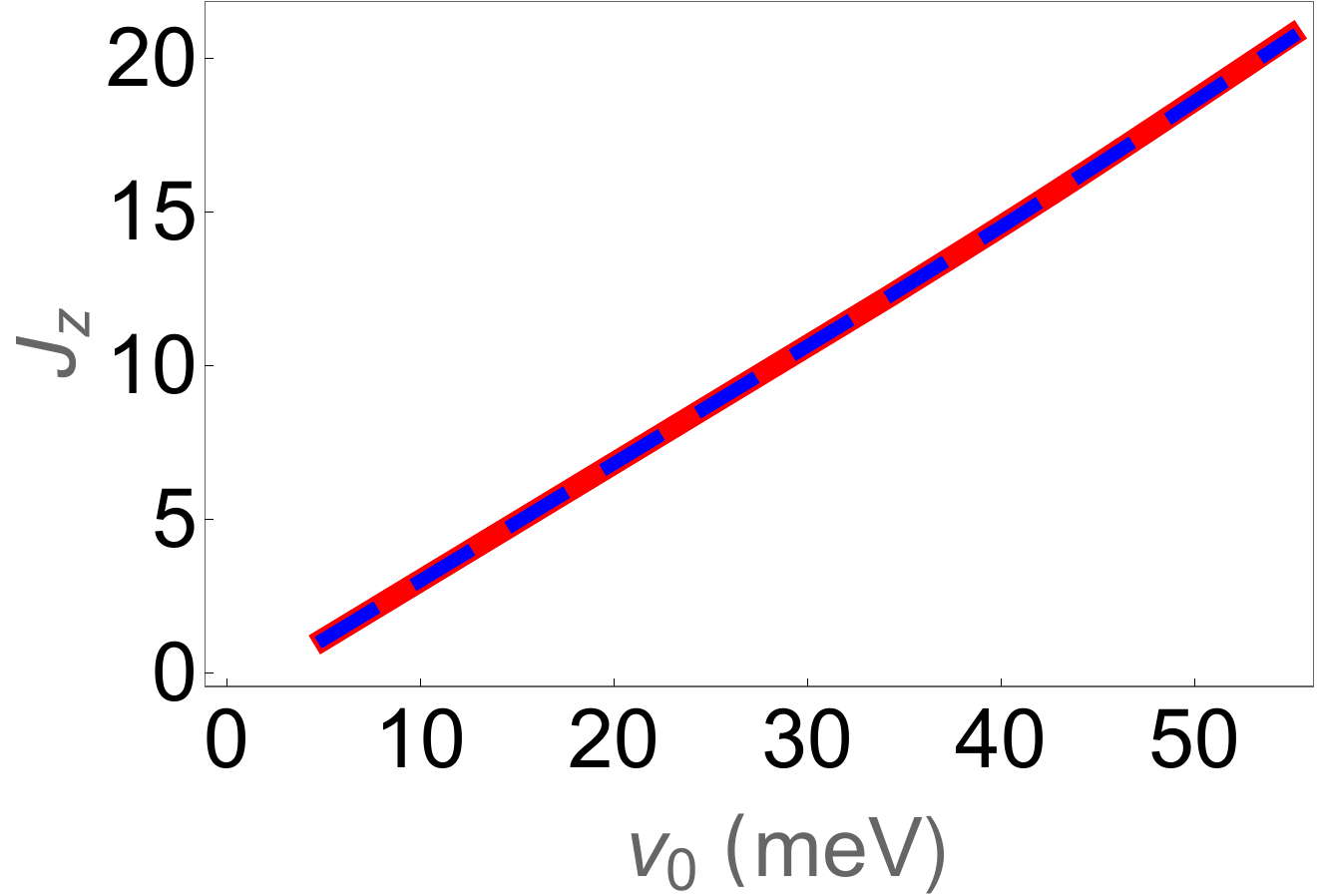}}
		\centerline{(c) }
	\end{minipage}
	\begin{minipage}{0.255\textwidth}		\centerline{\includegraphics[width=1\textwidth]{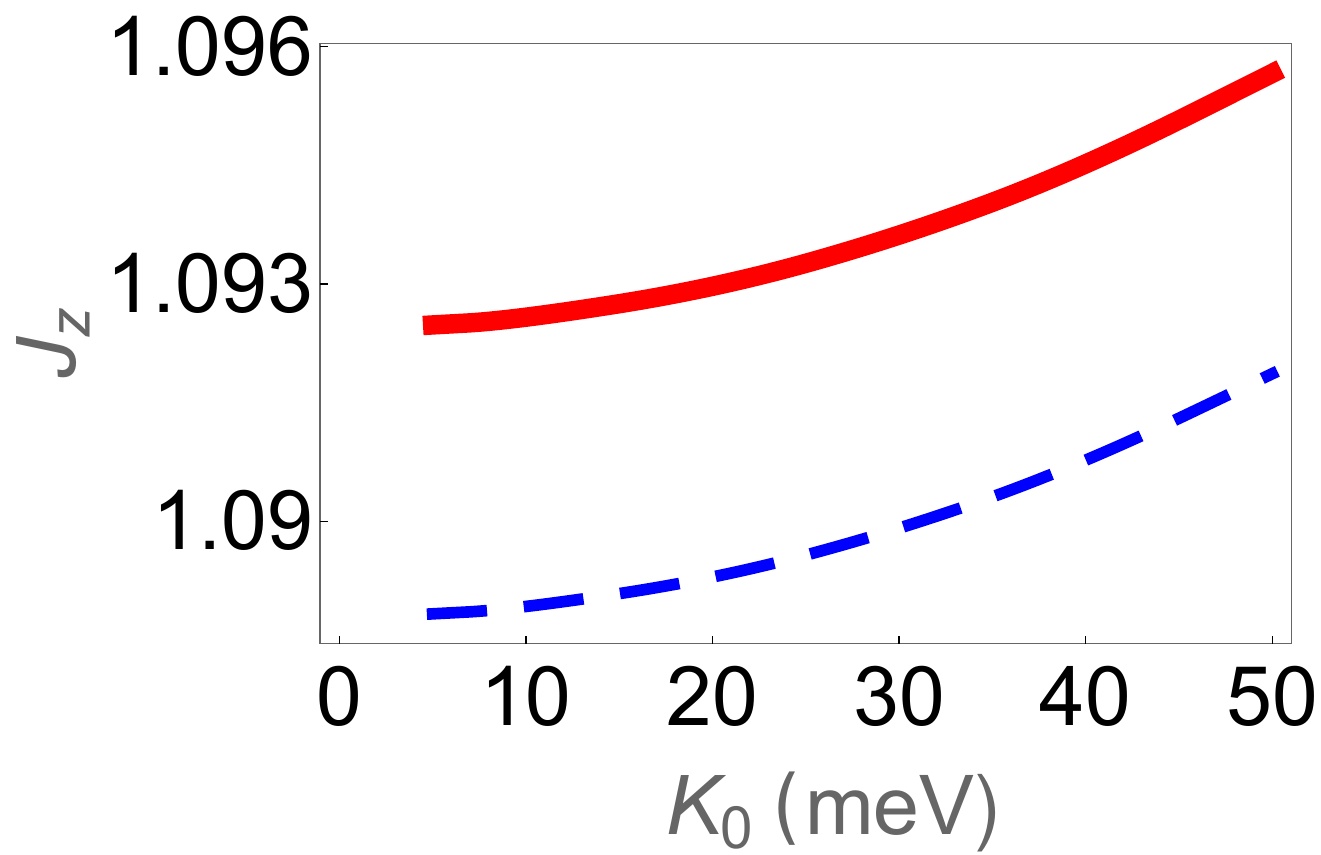}}
		\centerline{(d) }
	\end{minipage}
\caption{(color online)   {The current $J_z$ of the CME at zero temperature with $t=t'=0.1eV$,  $\phi=1/5$, and  $L_x=20$ is shown with periodic boundary conditions for the red solid lines and open boundary conditions for the blue dashed lines. $\Lambda=\pi/6$. (a) $K_0=10meV$, $Q=2\pi/3$;  {(b) $v_0=5meV$, $Q=2\pi/3$; (c) $K_0=10meV$, $Q=0$}; (d) $v_0=5meV$, $Q=0$.}}
	\label{fig3}	
\end{figure}

  To see the CME in Weyl semimetals, a cutoff $\Lambda\ll\pi $ in the summation of $k_z$ is taken instead of summing over the whole Brillouin zone and the current grows linearly in $b_0$ for a Weyl semimetal  \cite{Franz2013}. With this scenario, we calculate the CME with $L_x=20$, $\Lambda=\pi/6$, and plot $J_z$ by varying $v_0$ and $K_0$ at zero temperature in Fig. \ref{fig3}. We find that the Fermi arc states have a negative contribution to $J_z$ because $J_z$ under open boundary conditions is slightly smaller than that under periodic conditions. In the FM case, $J_z$ is linearly dependent on $v_0$ as expected  {(see Fig. \ref{fig3}c)}, while for the {CNHMC}, $J_z$ remains a constant when $v_0<3K_0$ and then grows (see Fig. \ref{fig3}a). For a large $v_0$, the folded Weyl nodes are separated far enough and isolated (see Fig. \ref{fig1}d). Therefore, the linear behavior of $J_z$ versus  $v_0$ is expected. In the $J_z$ constant regime, $J_z$  decreases when  $K_0$ increases. This shows that the Kondo coupling  suppresses the CME with the CNHMC  {(see Fig. \ref{fig3}b)} while it raises the CME with the FM order (see Fig. \ref{fig3}d). The key observation is that the CME in the CNHMC is about 20 times smaller than that in the FM order, which may explain the nearly zero behavior observed near $T=12.8K$ in CeAlGe \cite{Yu2023}.

\section{{Conclusions}} \label{secv}
We constructed a minimal lattice model with Kondo, FM, and DM interactions for the Weyl semimetal in which both $\Theta$ and $P$ symmetries are broken, which caught the key correlated physics in ReAlX.  We found that the CNHMC induces a magnetic Brillouin zone in the band structure of the conduction electrons. The magnetic phase transitions from the ordered phase to the CNHMC were explored. We also calculated the CME with the CNHMC, which is about 20 times smaller than that with the FM order. This may provide an explanation for the anomalous transport phenomena and nearly zero magnetoresistance observed in CeAlGe.    

 \acknowledgements

The authors thank Long Liang and Yong-Shi Wu for useful discussions. This work is supported by the National Natural Science Foundation of China with Grant No.~12174067 (XL, YMZ, and YY). 

\section*{DATA AVAILABILITY}

The data that support the findings of this article are not publicly available. The data are available from the authors upon reasonable request.

\newpage

\appendix

\section{Continuum limit and exactly soluble correlated helimagnet} \label{app1} 

When $|Q|\ll 2|k_0|$, the low energy effective Hamiltonian for the two Weyl nodes becomes
	\begin{equation}
		H_{eff}=\int d^3r \sum_{\nu=\pm}(-iv_\nu\psi^\dag_\nu{\boldsymbol\sigma}\cdot\nabla\psi_\nu+  {K_0}{\bf M}\cdot {\bf s_\nu}+  {\nu b_0})+...,
	\end{equation}
where $\nu$ labels the valley of the Weyl nodes, $v_\nu$ is the Fermi velocity of the Weyl node $\nu$, and   {$b_0$ is the energy offset between the two Weyl nodes caused by the inversion symmetry breaking.}  The "..." terms include the higher-order terms of derivatives and the exchange terms between ${\bf M}$s.   {Due to the energy offset $b_0$ between the two Weyl nodes in a non-centrosymmetric system}, we can integrate {out one Weyl fermion that is below the chemical potential,}  say,  {$\psi_-$}. 
The effective Hamiltonian is given by     {(the constant $b_0$ is omitted)} \cite{Yu16,Yu2023}
\begin{eqnarray}
	H_{eff}&=&\int d^3r\biggl[-i(\psi^\dag\boldsymbol\sigma\cdot\nabla\psi+K_0{\bf M}\cdot {\bf s})\nonumber\\
    &&+\frac{D}4{\bf M}\cdot{\nabla\times\bf M}+\frac{J}2(\nabla{\bf M})^2\biggr],\label{2}
\end{eqnarray}
{where the suffix + is omitted;   {$J\propto K_0^2-2J_0$} and   {$D\propto v_- K_0^2+4D_0$}} \cite{Xiao2015,rkky2}.  We have set $\hbar=v_+=1$ and neglected the high-order terms.
When $K_0=0$, the ground state configuration of the normalized ${\bf M}$  is an uncorrelated helimagnet, ${\bf M}=[{\bf e}_1\sin({\bf Q}\cdot{\bf r}+\varphi_0)\pm{\bf e}_2\cos({\bf Q}\cdot{\bf r}+\varphi_0) ]$
with $
   Q= |{\bf Q}|=|D/2J|$ and ${\bf Q}\propto {\bf e}_1\times {\bf e}_2 
$ \cite{Nagaosa2012}.
 When   {$K_0\neq 0$}, the spin ${\bf s}$ and ${\bf M}$   are coupled. The equation of motion for $\psi$ and the ground state configuration of ${\bf M}$ satisfy
 \cite{Yu16}
\begin{eqnarray}
	&&-i\boldsymbol\sigma\cdot(\nabla+i\frac{K_0}2{\bf M})\psi=E_e\psi, \label{effh} \\
	&&K_0{\bf s}+\frac{D}{2}\nabla\times {\bf M}-{J}\nabla^2{\bf M}=0. \label{effM}
\end{eqnarray}

Consider the following ansatz \cite{Don,freund,adam,sol1,sol2,sol3,3T}:
{
\begin{eqnarray}
     {\bf M}&=&(\sin[Qx], \cos[Qx],0),\\
     {\bf B}_0&=&(\pm)\frac{K_0}{2} \nabla\times{\bf M},\\
     \chi&=&\frac{e^{iQz/2}}{\sqrt{2(B_0+B_{03})}}\left(
    \centering
    \begin{tabular}{c}
      $B_0 + B_{03}  $ \\
      $B_{01}+iB_{02} $  \\
    \end{tabular}\right).
\end{eqnarray}}
One can check ${\bf B}_0=\chi^\dag {\boldsymbol \sigma} \chi$ and $B_0=\chi^\dag \chi$. Therefore this provides a map from $SU(2)$ to $SO(3)$. Then the equation of motion for $\psi=\chi/\sqrt{\chi^\dag \chi}$ reduces to 
\begin{equation}
    (\frac{Q}{2}+\frac{K_0}{2})\psi=E_e\psi,
\end{equation}
and the ground state equation for ${\bf M}$ reduces to  
\begin{equation}
     K_0{\bf M}+\frac{DQ}{2}{\bf M}-|J|Q^2{\bf M}=0.
\end{equation}
Therefore one obtains
\begin{equation}
    Q=\frac{D\pm \sqrt{D^2+16|J|K_0}}{4|J|}.
\end{equation}
  {We shall emphasize that although the helical configuration here has the same form as the case for $K_0=0$, $|{\bf Q}|$  is correlated with the Weyl fermions via the Kondo coupling. {It can be proved that the ground state solution can be found for a general helical magnetic configuration ${\bf M}=\frac{|Q|}{K_0}({\bf e}_1\sin({\bf Q\cdot r}+\phi_0)\pm {\bf e}_2\cos({\bf Q\cdot r}+\phi_0))$ \cite{Yu16}.}

  We present a more concrete solution with the ansatz ${\bf M}=\frac{Q}{K}(\sin[Qz],\cos[Qz],0)$. Assuming the periodic boundary condition, the spectrum $E$ of the fermions is determined by solving the equation of motion under the effective Hamiltonian (\ref{effh}): }
	\begin{equation}
		(-i\boldsymbol\sigma\cdot\nabla+\boldsymbol\sigma\cdot\frac{K}2{\bf M})\psi=E\psi.
	\end{equation}
	In order to find the zero-energy solution, we  focus along one direction, say, the $z$ axis, and take the following ansatz for the  wave function:
\begin{eqnarray}
	&&\psi_0({\bf r})=\psi_n(z)\nonumber\\
    &&\propto e^{i(k_z+nQ)z}\left(\begin{array}{c} 1\\ -\frac{2i}K(k_z+nQ-E)e^{-iQz}\end{array}\right), \label{ZM}
\end{eqnarray}  
{where $k_z$ is the wave vector along the $z$ axis, $Q$ is the magnetic wave vector of ${\bf M}$ , and $n$ labels the $n$th magnetic Brillouin zone determined by $Q$. Then, through a direct calculation, the spectrum reads
	\begin{equation}
		E_{\pm}=\frac{Q}2\pm\sqrt{(k_z+(n+\frac{1}2)Q)^2+\frac{K^2}4}. \label{spec}
	\end{equation}
	The zero energy occurs when $|Q|=|K|$ (which is consistent with the normalization $|{\bf M}|=1$),  and $k_z=-(n+\frac{1}2)Q$.}  {The diagram of the spectra (\ref{spec}) is shown in Fig. \ref{fig0} of the main text.} The spin of these zero modes is the same, i.e.,  ${\bf s}_n$=${\bf s}= \frac{1}2\frac{Q}K{\bf M}$.  Eq. (\ref{effM}) determines $Q$ as  %{We plot a diagram of the spectra (\ref{spec}) in Fig. \ref{sup0}.} 
    
\begin{figure}
	\includegraphics[width=0.23\textwidth]{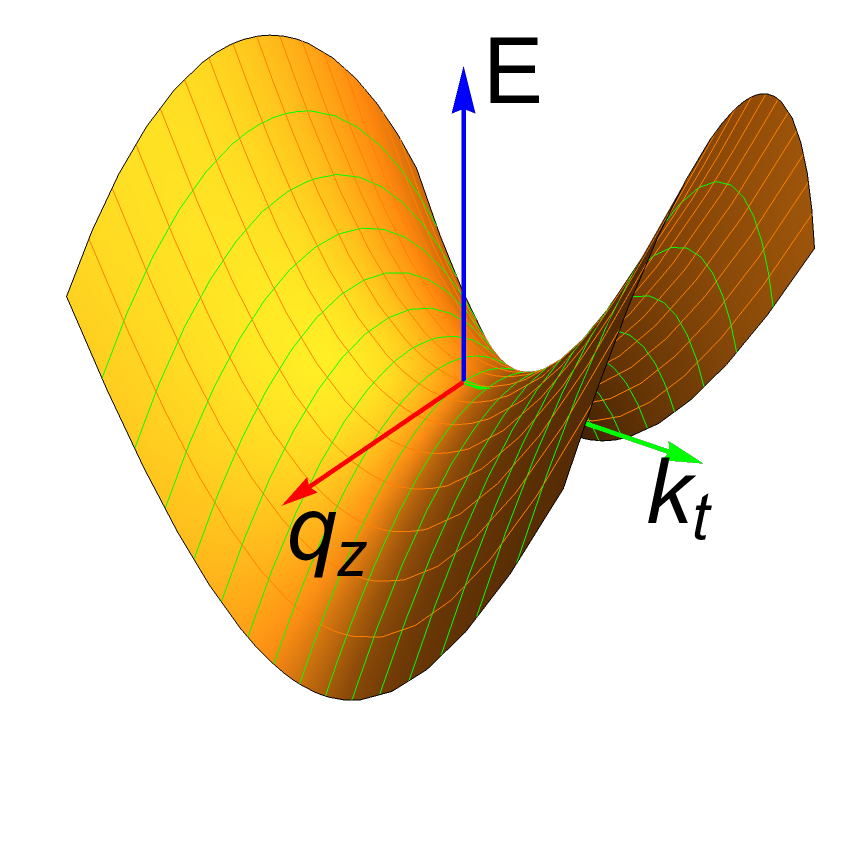}
\caption{(color online)    {A schematic diagram for the dispersion of $E_-$ (\ref{eq12}) near the blue points in Fig. \ref{fig0}. $K=1$. The dispersion has a saddle shape. It is a parabola along $k_t$ and a  downward one along $q_z$. }}
	\label{sup1}	
\end{figure}

   % \begin{figure}
%	\includegraphics[width=0.3\textwidth]{fig0.pdf}
	%	\includegraphics[width=3.4in]{fig1.eps}
%\caption{(color online)    {The band structure of Eq. (\ref{spec}) with different $n$s. $Q=K=1$. The magnetic Brillouin zone is introduced by the helical magnetic order. The red points stand for the Weyl nodes and the blue points stand for the  zero modes.}}
%	\label{sup0}	
%\end{figure}

\begin{eqnarray}
	\frac{Q}{2K}+\frac{DQ}{2K}-\frac{|J|Q^2}K=0\Rightarrow Q=\frac{1}{2|J|}(1+D), 
\end{eqnarray}\\
which also gives a constraint between $K$ and $J/D$. %This may explain the 12.8K phase can only appear in a very narrow temperature region. 
For a pinned helimagnet with $Q>0$, say, in the $z$ direction, the dispersion  near  {$k_z=-(n+\frac{1}2)Q$} for a given zero mode reads,
\begin{eqnarray}  
	E_{-}(k_z)\approx -\frac1{|K|} (k_z+(n+\frac{1}2)Q)^2\equiv -\frac1{|K|} q_z^2. 
\end{eqnarray}
We now perform  the perturbation by taking $H_1=k_x\sigma^x+k_y\sigma^y$ because for the pinned helimagnet, the Hamiltonian  (\ref{2}) in the main text is translationally invariant in the $x$-$y$ plane. The first-order perturbation is zero. The matrix elements between
\begin{eqnarray}
   && \psi_{n-}(z)\propto e^{i(k_z+nQ)z}\left(\begin{array}{c} 1\\ -\frac{i2}K(k_z+nQ-E_{n-})e^{-iQz}\end{array}\right)\nonumber\\
   &=&e^{i(k_z+nQ)z}\left(\begin{array}{c} 1\\ A_ne^{-iQz}\end{array}\right) \nonumber
\end{eqnarray}
  and $\psi_{n'-}(z)$ are 
\begin{eqnarray}
	&&\langle \psi_{n'-}(z)|H_1|\psi_{n-}(z)\rangle\nonumber\\
    &=&\delta_{n,n'+1}A_n(k_x-ik_y)+\delta_{n,n'-1}A_{n'}(k_x+ik_y).
    \end{eqnarray}
The energy differences  at $k_z=-(n+\frac{1}2)Q$ are
\begin{eqnarray}
	E_{-}(n,k_z)-E_{-}(n\pm1,k_z)=\frac{\sqrt5-1}2K,
\end{eqnarray}
The second order perturbation gives
\begin{eqnarray}
	E^{(2)}_{-}&=&\sum_{n'\ne n}\frac{\langle \psi_{n-}(z)|H_1|\psi_{n'-}(z)\rangle \langle \psi_{n'-}(z)|H_1|\psi_{n-}(z)\rangle}{E_n-E_{n'}}\nonumber\\
	&=&\frac{2}{\sqrt5-1}\frac{k_x^2+k_y^2}K.
\end{eqnarray}
Thus, the dispersions near $k_z=-(n+1/2)Q$ are 
\begin{eqnarray}
	E_{-}({\bf k})&=&-\frac{1}K q_z^2+\frac{2}{(\sqrt5-1)K}k_t^2\nonumber\\
    &\equiv&-\frac{q_z^2}{2m_z}+\frac{k_t^2}{2m_t}=-\varepsilon_z+\varepsilon_t,\label{eq12}
    \end{eqnarray}
with ${\bf k} _t=(k_x,k_y)$. 
The dispersion {is of a saddle shape   {(see Fig. \ref{sup1})}; namely,} in the $z$ direction, it is hole-type, while in the $x$-$y$ plane, it is particle-type.  This means that the dynamics  parallel to the helimagnet and  perpendicular to the helimagnet are separated.

\section{A brief review of the two-dimensional lattice model in a constant external magnetic field and the Harper equation} \label{app2}

Consider a tight-binding Hamiltonian on the square lattice in a magnetic field \cite{Kohmoto1989,Kohmoto1990,Hatsugai1993}:
\begin{eqnarray}
    H&=&-t_a\sum_{\langle i,j\rangle}c_{x,j}^\dagger c_{x,i}e^{i\theta^x_{ij}}\nonumber\\
    &&-t_b\sum_{\langle i,j\rangle}c_{y,j}^\dagger c_{y,i}e^{i\theta^y_{ij}}+h.c., 
\end{eqnarray}
where $\theta_{ij}$ is the phase factor of the gauge field which satisfies $\sum_\boxdot \theta_{ij}=2\pi \phi$ in a unit cell, and $\phi=p/q$ with $p,q$ coprime. $t_a$ and $t_b$ are the nearest-neighbor hopping constants.  A site $i$ on the square lattice has  Cartesian coordinates $(n, m)$ where $n$ and $m$ are integers. We choose the gauge $\theta^x_{ij}=0$, and $\theta^y_{ij}=2\pi \phi n$ for the link between $i=(n,m)$ and $j=(n,m+1)$. This gauge gives a
uniform magnetic field the flux of which through a plaquette is $2\pi \phi$. In the momentum space, 
\begin{eqnarray}
    H({\bf k})&=&-2t_a \cos k_x c^\dag(k_x,k_y)c(k_x,k_y)\nonumber\\
    &&-t_b[e^{-ik_y}c^\dag(k_x+2\pi \phi,k_y )c(k_x,k_y)\nonumber\\
    &&+e^{ik_y}c^\dag(k_x-2\pi \phi,k_y )c(k_x,k_y)],    
\end{eqnarray}
where $-\pi\leq k_x,k_y\leq \pi$. Since $k_x$ is coupled to $k_x\pm2\pi \phi$, the Schr\"odinger equation $H|\Psi\rangle=E|\psi\rangle$ is reduced to the Harper equation:
\begin{eqnarray}
    &&-t_b (e^{-ik_y}\psi_{j-1}+e^{ik_y}\psi_{j+1})-2t_a\cos (k^0_x+2\pi\phi j)\psi_j\nonumber\\
    &&=E(k_x^0,k_y)\psi_j, \label{Harper}
\end{eqnarray}
where $k_x=k_x^0+2\pi \phi j$ and $\phi_{j+q}=\phi_j$. And the state $|\Psi\rangle$ is
\begin{equation}
    |\Psi\rangle=\sum_{j=1}^q\psi_jc^\dag (k_x^0+2\pi\phi j,k_y)|0\rangle.
\end{equation}
%When $\phi=p/q$ is a rational number, the Harper equation has $q$ eigenvalues for a given $(k^0_x,k_y)$. The original band for the tight-binding model is split into $q$ bands due to the application of the magnetic field, and each band has a reduced magnetic Brillouin zone, $-\pi/q \leq k_x^0\leq \pi/q$, $-\pi\leq k_y\leq \pi$. 

\section{Evolution of Weyl nodes with respect to the Kondo coupling $K_0$} \label{app3}

\begin{figure}
	%\begin{minipage}{0.23\textwidth}	\centerline{\includegraphics[width=1\textwidth]{sup3a.pdf}}
%		\centerline{(a) }
%	\end{minipage}
%	\begin{minipage}{0.23\textwidth}		\centerline{\includegraphics[width=1\textwidth]{sup3b.pdf}}
%		\centerline{(b) }
%	\end{minipage}
    \begin{minipage}{0.23\textwidth}	\centerline{\includegraphics[width=1\textwidth]{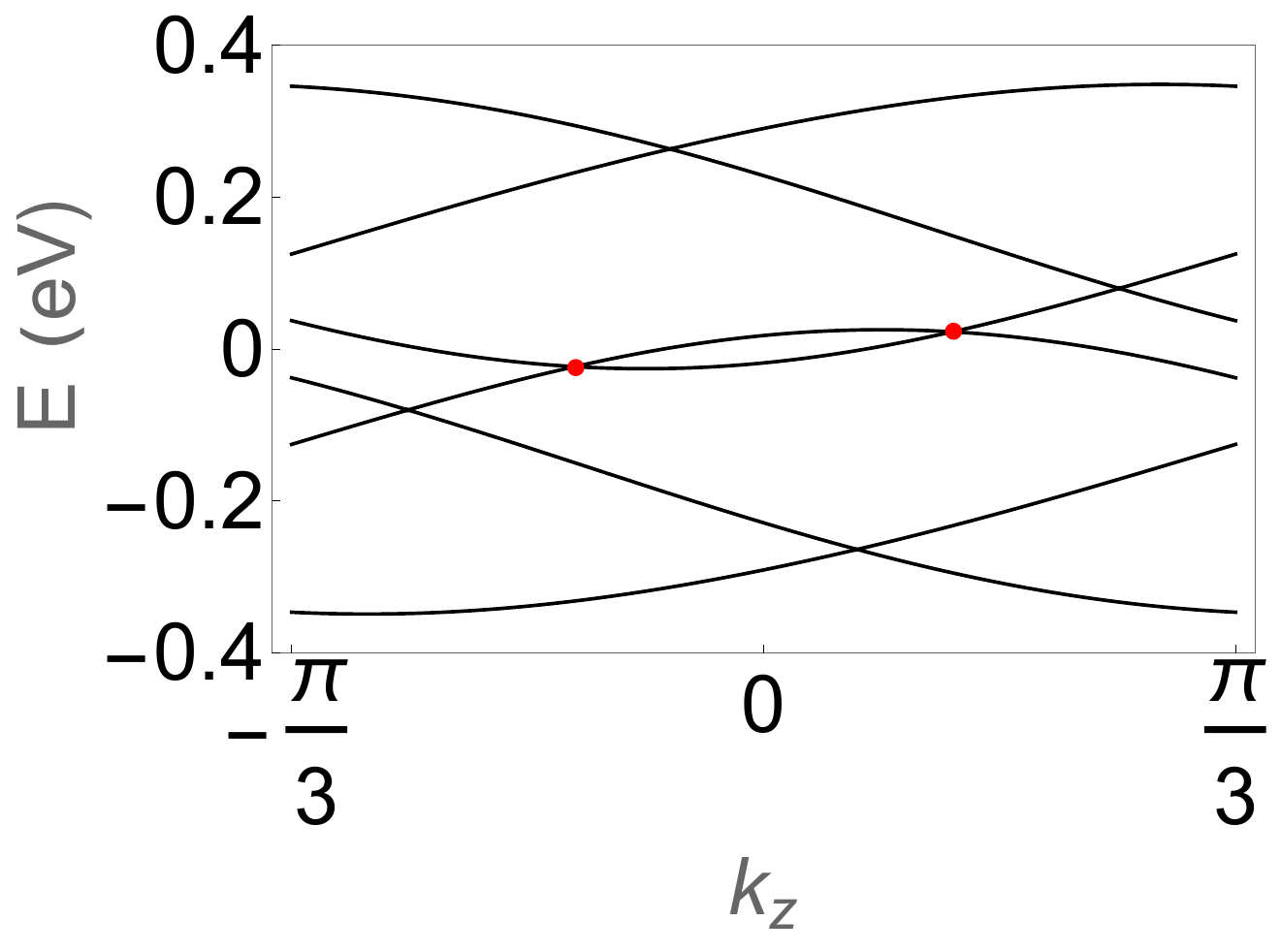}}
		\centerline{(a) }
	\end{minipage}
	\begin{minipage}{0.23\textwidth}		\centerline{\includegraphics[width=1\textwidth]{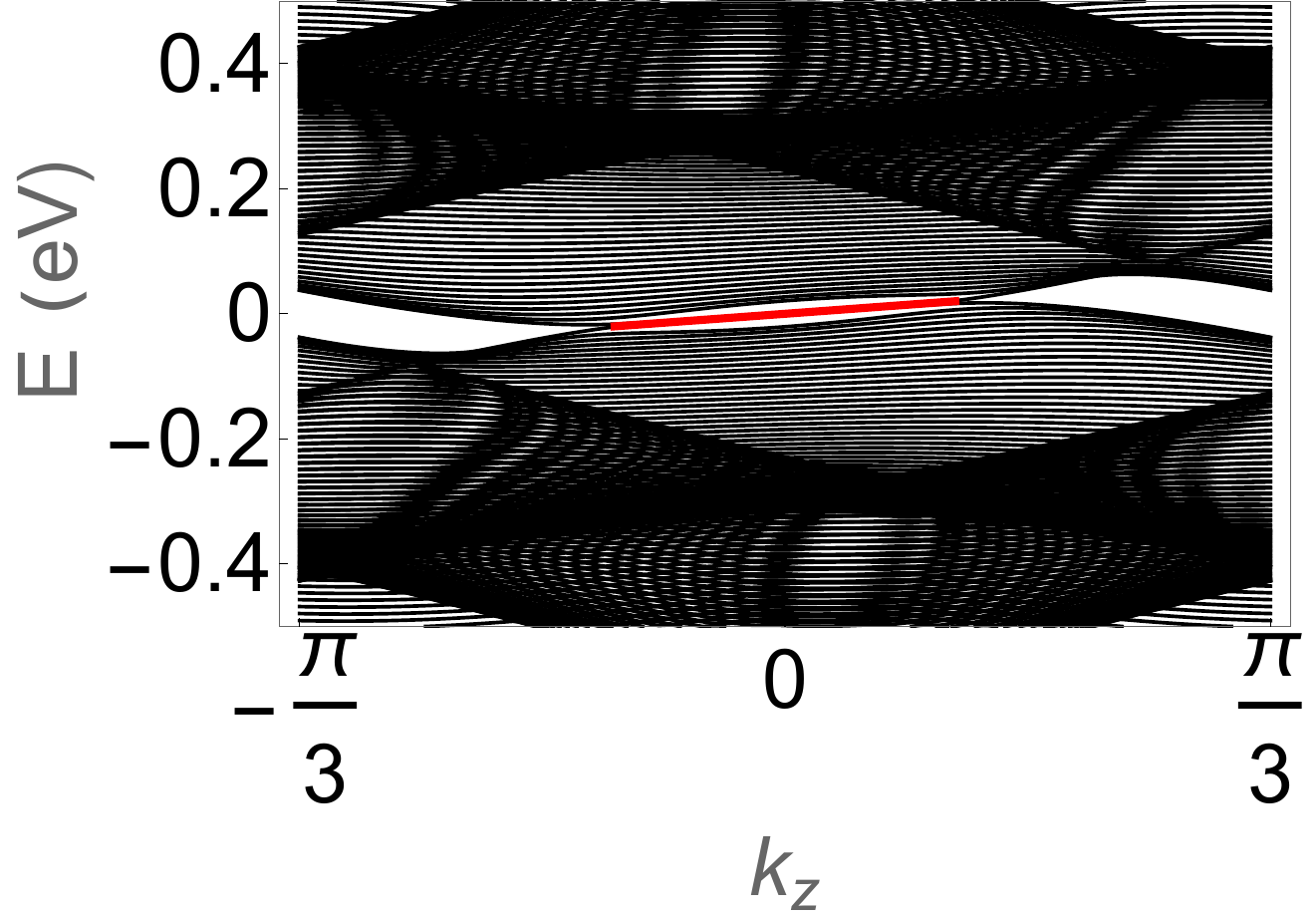}}
		\centerline{(b) }
	\end{minipage}
    	\begin{minipage}{0.23\textwidth}	\centerline{\includegraphics[width=1\textwidth]{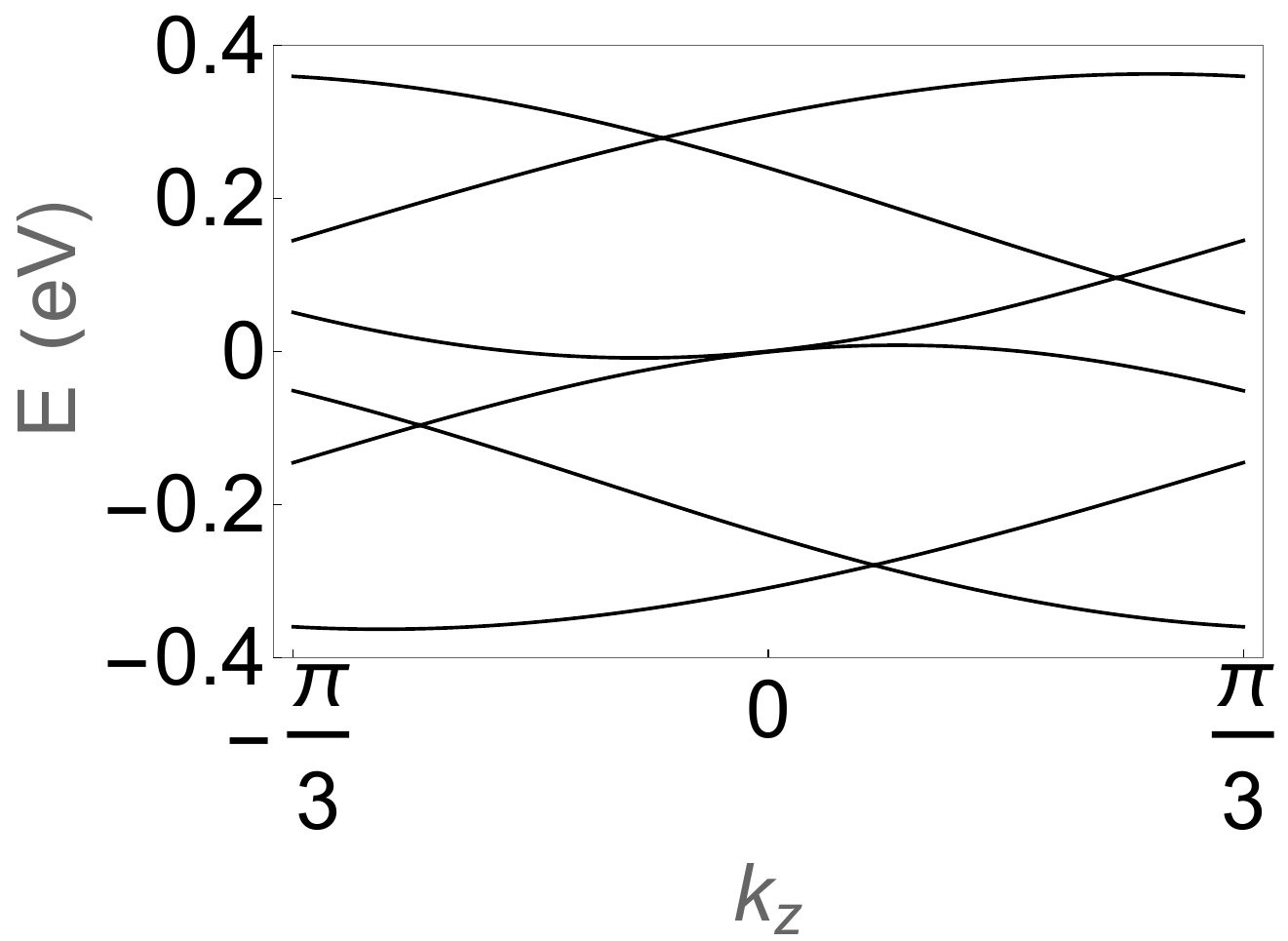}}
		\centerline{(c) }
	\end{minipage}
    	\begin{minipage}{0.23\textwidth}	\centerline{\includegraphics[width=1\textwidth]{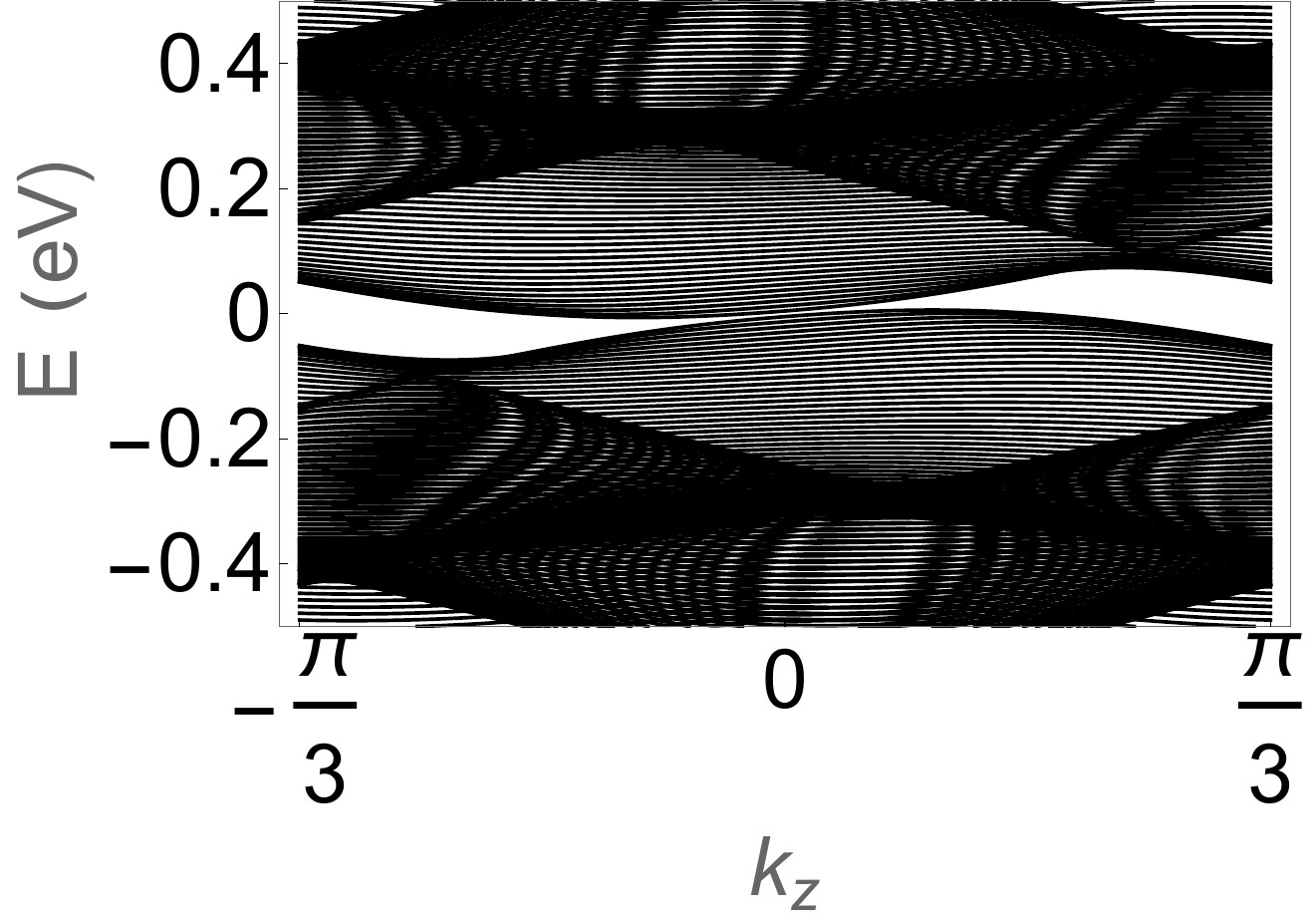}}
		\centerline{(d) }
	\end{minipage}
    	\begin{minipage}{0.23\textwidth}	\centerline{\includegraphics[width=1\textwidth]{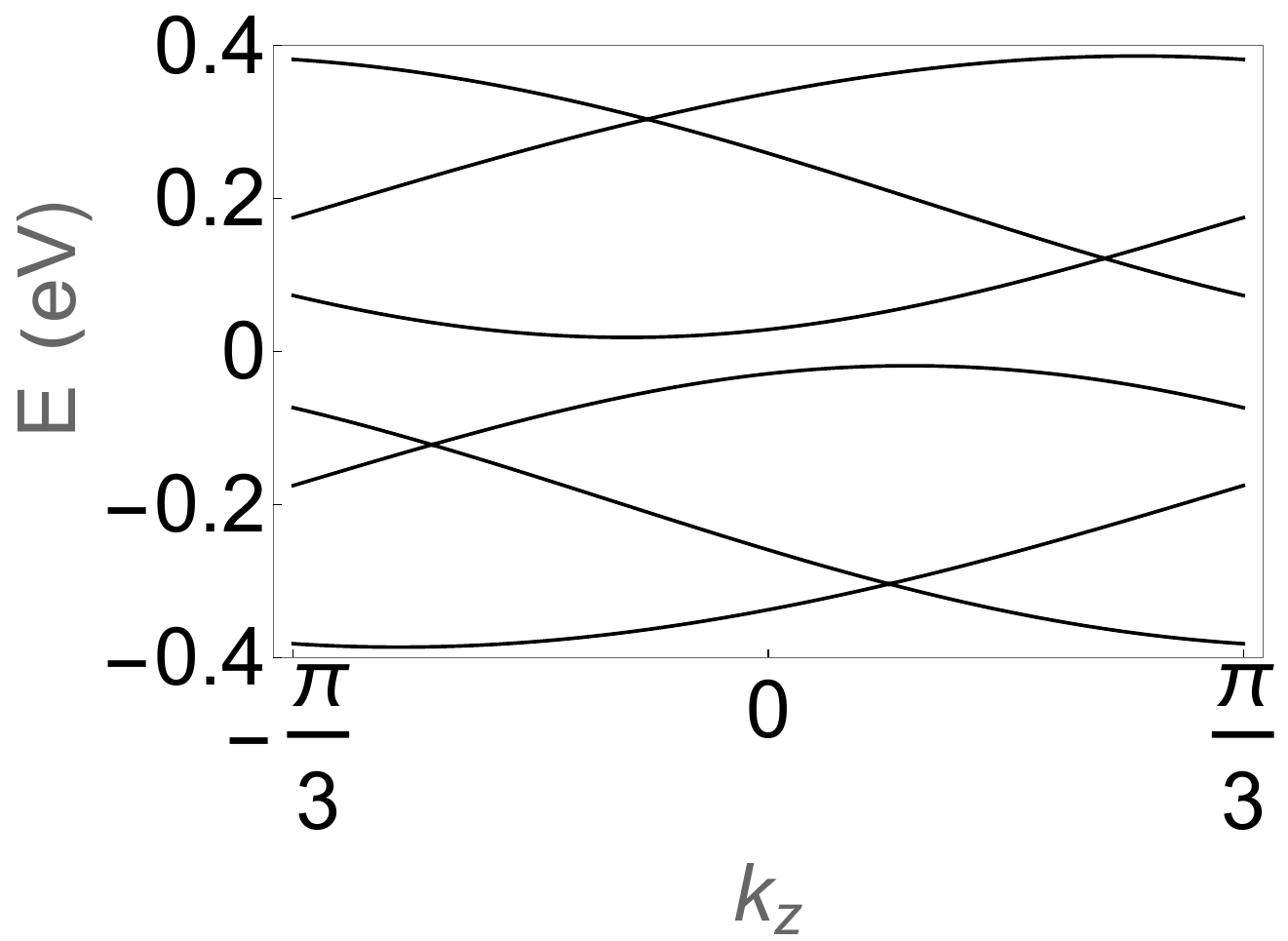}}
		\centerline{(e) }
	\end{minipage}
    	\begin{minipage}{0.23\textwidth}	\centerline{\includegraphics[width=1\textwidth]{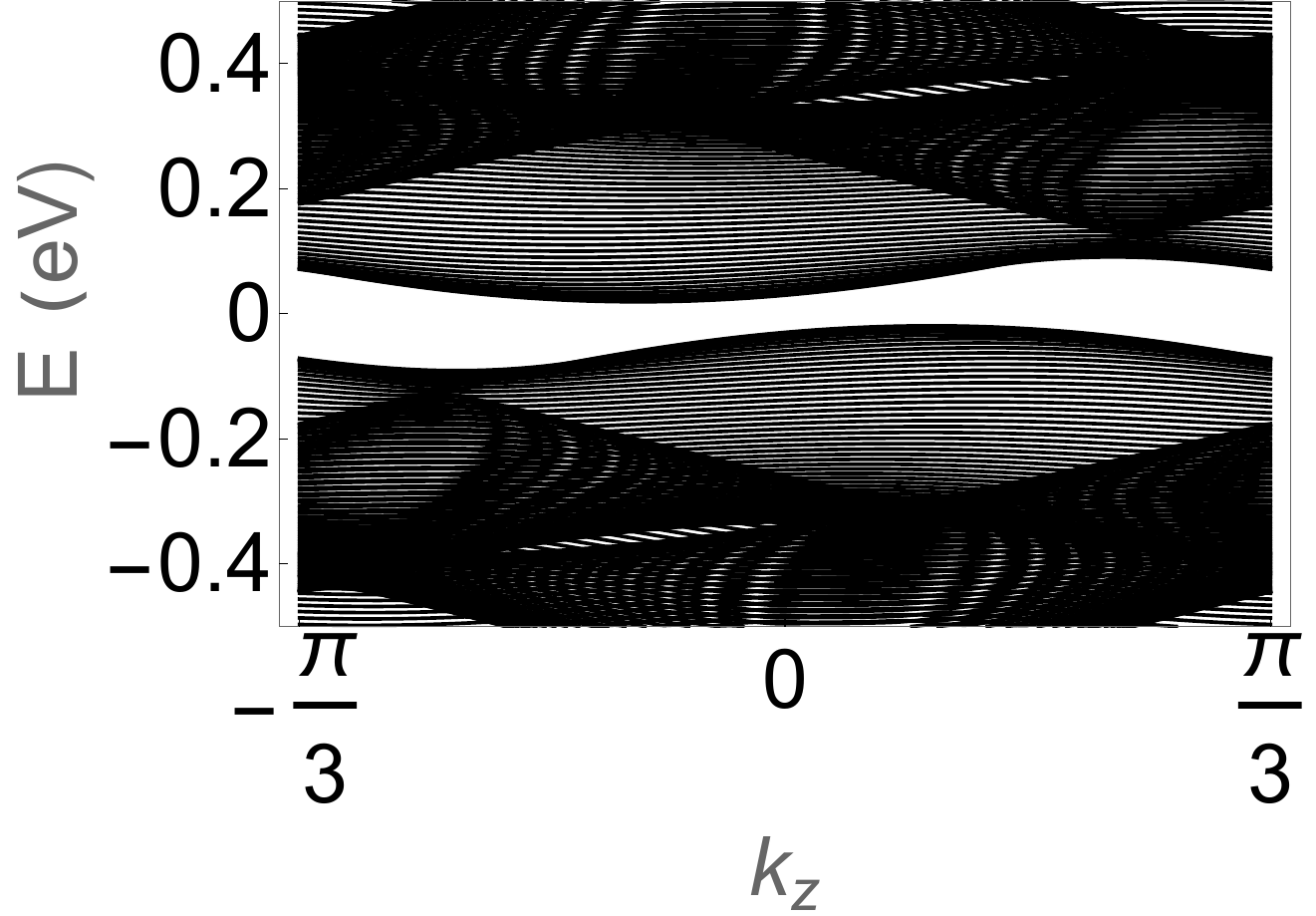}}
		\centerline{(f) }
	\end{minipage}
\caption{(Color online)  The evolution of Weyl nodes. The parameters are chosen as $Q=2\pi/3$, $t=t'=0.1eV$, and $k_x=k_y=0$.  (a) $K_0=250meV$, (c) $K_0=290meV$, (e) $K_0=350meV$. (b), (d), and (f) are the corresponding band structures with open boundary conditions for 50 sites along the $y$-direction. The red points are Weyl nodes, and the red lines are the Fermi arcs.}
	\label{sup3}	
\end{figure}

We plot the evolution of Weyl nodes with respect to the  Kondo coupling $K_0$ in Fig. \ref{sup3}. We find that by increasing the Kondo coupling, the Weyl nodes  move closer to each other. Finally, they annihilate and open a gap. This process reflects the competition between the monopole topology of the Weyl node and the trivial topology of the helimagnet via Kondo coupling $K_0$. For a small $K_0$, the Weyl node structure remains stable, while for a large $K_0$, the conducting fermions acquire a gap and become trivial.

\end{document}